\documentclass[submission, Phys]{SciPost}

\usepackage[utf8]{inputenc}
\usepackage[english]{babel}
\usepackage[T1]{fontenc}
\usepackage{amsmath}
\usepackage{hyperref}
\usepackage{bm}
\usepackage{environ}
\usepackage{graphicx}
\usepackage{amssymb,amsmath}
\usepackage{bm}
\usepackage{dcolumn}
\usepackage{float}
\usepackage[OT1]{fontenc} 
\usepackage{url}
\usepackage{slashed}
\usepackage{color}
\usepackage{verbatim}
\usepackage{txfonts}
\usepackage{soul}
\usepackage{comment}
\usepackage{amssymb}

\usepackage{lipsum}

\hypersetup{colorlinks = true, linkcolor=black, citecolor=black, urlcolor=blue}
\usepackage{url}
\usepackage{tikz,fp}
\usepackage{tikz-cd}
\usetikzlibrary{arrows}
\usetikzlibrary{intersections}
\usetikzlibrary{shapes.geometric}
\usetikzlibrary{decorations.pathmorphing, patterns,shapes,fixedpointarithmetic}
\usetikzlibrary{decorations.markings}

\newcommand{\RN}[1]{%
	\textup{\uppercase\expandafter{\romannumeral#1}}%
}
\newcommand{\Rn}[1]{%
	\textup{\lowercase\expandafter{\romannumeral#1}}%
}

\newcommand*\diff{\mathop{}\!\mathrm{d}}

\pgfdeclarepatternformonly{south west lines}{\pgfqpoint{-0pt}{-0pt}}{\pgfqpoint{3pt}{3pt}}{\pgfqpoint{3pt}{3pt}}{
  \pgfsetlinewidth{0.4pt}
  \pgfpathmoveto{\pgfqpoint{0pt}{0pt}}
  \pgfpathlineto{\pgfqpoint{3pt}{3pt}}
  \pgfpathmoveto{\pgfqpoint{2.8pt}{-.2pt}}
  \pgfpathlineto{\pgfqpoint{3.2pt}{.2pt}}
  \pgfpathmoveto{\pgfqpoint{-.2pt}{2.8pt}}
  \pgfpathlineto{\pgfqpoint{.2pt}{3.2pt}}
  \pgfusepath{stroke}}

\tikzset{
  mid arrow/.style={postaction={decorate,decoration={
        markings,
        mark=at position .575 with {\arrow{stealth}}
  }}},
  near arrow/.style={postaction={decorate,decoration={
        markings,
        mark=at position .275 with {\arrow{stealth}}
  }}},
  far arrow/.style={postaction={decorate,decoration={
        markings,
        mark=at position .800 with {\arrow{stealth}}
  }}},
  snake arrow/.style={fixed point arithmetic, decorate, decoration={snake,amplitude=2pt, segment length=11pt},postaction={decoration={markings,mark=at position 0.575 with {\arrow{stealth}}},decorate}},
}

\tikzset{
  baseline = -0.5ex,
  wavy/.style = {
    thick,
    decorate,
    decoration={snake,amplitude=2pt,segment length=5pt}},
  sdot/.style = {
    circle,
    draw=none,
    fill=black,
    minimum size=2.5pt,
    inner sep=0pt},
  bdot/.style = {
    circle,
    draw=none,
    fill=black,
    minimum size=4pt,
    inner sep=0pt},
  svertex/.style = {
    circle,
    draw=black,
    thick,
    fill=lightgray,
    minimum size=8pt,
    inner sep=1pt},
  bvertex/.style = {
    circle,
    draw=black,
    thick,
    fill=lightgray,
    minimum size=16pt}}

\begin{document}

\begin{center}{\Large \textbf{ Generalized Lindblad Master Equation for Measurement-Induced Phase Transition
}}\end{center}

\begin{center}
Yi-Neng Zhou
\end{center}

\begin{center}
Institute for Advanced Study, Tsinghua University, Beijing,100084, China

* zhou-yn19@mails.tsinghua.edu.cn
\end{center}

\begin{center}
\today
\end{center}


\section*{Abstract}
{\bf The measurement-induced phase transition (MIPT) occurs when the system is evolving under unitary evolution together with local measurements followed by post-selection. We propose a generalized version of the Lindblad master equation as a continuous equation, to describe the dynamics of the second R\'enyi entropy in the MIPT. This generalized Lindblad equation explicitly takes into account the post-selection in the MIPT, which is realized as the Einstein-Podolsky-Rosen (EPR) state projection in the equation. Also, this generalized Lindblad equation preserves the Hermitian, unit trace, and positive definiteness of the density matrix. We further use the hard-core Bose-Hubbard model as a concrete example to numerically confirm that our generalized Lindblad equation is applicable to describing the MIPT. 
}

\vspace{10pt}
\noindent\rule{\textwidth}{1pt}
\tableofcontents\thispagestyle{fancy}
\noindent\rule{\textwidth}{1pt}
\vspace{10pt}

\section{Introduction}
Recent years have seen major progress in the understanding
of quantum entanglement, and the study of entanglement
transitions that separate different entanglement phases have been wildly discussed. Since the unitary evolution of a closed system typically drives it towards volume-law scaling for the entanglement entropy of subsystems, adding local measurement in the evolution process has been raised as a method to restrict the growth of entanglement. The study of the entanglement transition gave rise to the concept of the measurement-induced phase transition (MIPT).  In measurement-induced phase transition, the unitary evolution can establish the entanglement between different parts of the system, while the local projective measurements on the system are believed to destroy the entanglement between different parts of the system. Thus, there is a competing relation between the unitary evolution and local measurements, and therefore, the increase in measurement rate can lead to an entanglement phase transition   \cite{entanglementFisher2018,entanglementFisher2019,entanglementNahum2019,entanglement_weak,entanglementAmosChan2019,ultracold_experiment,quantum_measure}. This dynamical process in some cases can be mapped to a classical percolation problem, making it feasible for large-size classical simulation \cite{percolation}. It also enables the study of the phase diagram and critical exponents in different regions \cite{critical_exponent_Random_circuit,critical_exponent_Fermion,critical_exponent_Fermion2,critical_exponent_quantum_trajectory,phase_diagram,critical_random circuit,critical_circuit_3Dpercolation}. The nature of this phase transition has been analyzed from different perspectives including classical statistical mechanics models \cite{percolation,statisticRUC}, information scrambling \cite{purification}, quantum error corrections \cite{quantum_error,quantum_error2021} and symmetry breaking \cite{syk_symmetry}. 

The methods of studying entropy dynamics under measurements include tensor network \cite{RTN,tensor_network,all_to_all_circuit}, matrix product state \cite{MPS}, random unitary circuit \cite{RUC,statisticRUC}. However, it remains an open question to find an equation that is continuous in time to describe the entropy dynamics of a quantum system along this unitary evolution together with measurements. Also, the role of post-selection is essential in this entanglement phase transition since it excludes the entropy corresponding to the probability distribution of different measurement results, and thus makes it possible to obtain the entanglement transition. Therefore, it should be explicitly written in this continuity equation. Here, we consider the dynamics of the second R\'enyi entropy of the system under unitary evolution together with projective measurements, followed by post-selections to project a general mixed state to a pure state. 

In this paper, we derive a dynamical equation of density matrix to describe the second R\'enyi entropy dynamics under unitary evolution and projective measurements that are followed by post-selections. This generalized Lindblad equation preserves the Hermitian, unit trace, and positive definiteness of the density matrix. Moreover, in this process, the entropy comes from two parts: the entanglement entropy of the system corresponding to different measurement results, and the entropy corresponding to the probability distribution of different measurement results. In the MIPT, we only care about the entropy of the former part, so we need to exclude the entropy of the latter part.  In our equation, we use the measurement on a partial basis to directly exclude this part of entropy. Since the measurement on the partial basis is explicitly written in our equation, we can observe the entanglement phase transition by directly calculating the entanglement entropy from our equation. Thus, we do not need to run the same protocol many times and post-process the entropy results to observe this phase transition. Our generalized Lindblad equation thus provides a natural description of the entropy dynamics along this process.

\section{The generalized Lindblad equation for measurement process \label{measurement_Lindblad}}

We first briefly review how to obtain a Lindblad-like equation that describes a system under unitary evolution together with frequent measurements \cite{measurement_book,master_equation,measurement_add,open_system}.  
\\For a closed system under unitary evolution, the time evolution of the density matrix of the system follows:
\begin{equation}
	\frac{\partial \rho}{\partial t} =-i[\hat{H},\rho].
\end{equation} 
Thus, to the first order of $\delta t$, we obtain
\begin{equation}
	\rho(t+\delta t)\simeq \rho(t)-i[\hat{H},\rho(t)]\delta t+o(\delta t)^{2}.
\end{equation} 
Next,  we consider the system being measured at an equal time interval $\delta t$. During two neighboring measurements, the system's evolution is governed by $\hat{H}$. Then, the density matrix after the measurements is
\begin{equation}
	\rho^{M}(t+\delta t)=\sum_{a}\hat{L}_a\rho(t+\delta t)\hat{L}_a^{\dagger}.
\end{equation} 
Here, we use $\rho^{M}$ to denote the density matrix after measurements $(M)$, and $\hat{L}_a$ is called the Lindblad operator or quantum jump operator. Here, $a=1,...,n$ labels the possible quantum jumps resulting from measurements.

If we simply assume that the probability of the system being measured at time $t+\delta t$ is $P(t+\delta t )$,
then the density matrix after this probabilistic measurement is:
\begin{equation}
	\label{measurement_process}
		\rho^{M}(t+\delta t)=\left[1-P(t+\delta t)\right]\rho(t+\delta t)+P(t+\delta t)\sum_{a}\hat{L}_a\rho(t+\delta t)\hat{L}_a^{\dagger}.
\end{equation} 
Here, we consider the system is projected to a complete basis 
\begin{equation}
	\label{complete}
	\sum_{a=1}^{n}\hat{L}_a^{\dagger}\hat{L}_a=\mathcal{I},
\end{equation}
and this assumption preserves the trace of density matrix along this measurement process.
By  assuming the completeness of measurement basis and taking the limit $\delta t \to 0$ then we obtain a differential equation of density matrix
\begin{equation}
		\label{without_post}
		\frac{\partial \rho}{\partial t}=-i\left[H,\rho(t)\right]+\eta(t)\sum_{a=1}^n\left[\hat{L}_a\rho(t )\hat{L}_a^{\dagger}-\frac{1}{2}\lbrace \hat{L}_a^{\dagger}\hat{L}_a,\rho(t)\rbrace\right].
\end{equation}
Here, $\eta(t)$ is the probability of the system being measured per unit of time, and it is defined as $\eta(t+\delta t)=\frac{P(t+\delta t)}{\delta t}$. Here, we suppose that there is no singularity in $\eta(t)$. This equation is the same as the Lindblad master equation if we regard the measurement rate $\eta(t) $ as the dissipation strength $\gamma$.
\\Also, we mention that here we assume the measurement is performed on a complete basis, thus the anti-commutator part of the Lindblad master equation is trivial and amounts to an identity matrix, therefore, the equation is a special case of the general Lindblad equation. Also, the system under unitary evolution together with the measurement process can be described in other ways.   For instance, we can derive the Lindblad master equation from a weak measurement approach \cite{weak_measurement_1988,weak_measurement_2021,critical_exponent_Fermion2,quantum_Zeno_2020}, which is by coupling the system to an ancilla and then performing a projective measurement on it and use the operator-sum representation, then $	\sum_{a=1}^{n}\hat{L}_a^{\dagger}\hat{L}_a\neq\mathcal{I}$ in general. We put more discussion about the the assumption $	\sum_{a=1}^{n}\hat{L}_a^{\dagger}\hat{L}_a=\mathcal{I}$ in the Eq.~(\ref{without_post}) and the “weak measurement” scenario where $	\sum_{a=1}^{n}\hat{L}_a^{\dagger}\hat{L}_a\neq \mathcal{I}$ in the section \ref{difference_section} of the supplementary\cite{Supplementary}.

\section{The measurement process followed by post-selection}

When one considers the case where the system is projected on a partial basis, the Eq.~(\ref{without_post}) is not able to describe the density matrix dynamics. Here, projection on a partial basis means we only save the result of the system being projected on some specific states $b$ where $b=1,2,...,m$ and $m<n$. To see  the failure of the Eq.~(\ref*{without_post}) in this case, we can first directly use the Eq.~(\ref{without_post}) by changing the summation on the right-hand side from $a=1,...,n$ to $a=1,...,m$. Then, we will find that this new equation does not preserve the trace of the density matrix. Therefore, to preserve the unit trace of the density matrix, we need to normalize the density matrix when the system has been measured. Thus, in comparison to the Eq.~(\ref{measurement_process}),  after this probabilistic measurement and the followed post-selection, the density matrix becomes:
\begin{equation}
		\rho^{M}(t+\delta t)=\left[1-P(t+\delta t)\right]\rho(t+\delta t) +P(t+\delta t)\frac{\sum_{a=1}^m\hat{L}_a\rho(t+\delta t)\hat{L}_a^{\dagger}}{\mathrm{Tr}\left(\sum_{b=1}^m\hat{L}_b\rho(t)\hat{L}_b^{\dagger}\right)}.
\end{equation} 
Here, the denominator on the right-hand side is the normalization factor due to post-selection. 
Similar to the Eq.~(\ref{without_post}), by taking the limit $\delta t \to 0$, we further obtain:
\begin{equation}
	\label{post_selection_partial}
		\frac{\partial \rho}{\partial t}=-i[\hat{H},\rho(t)]+\eta(t)\sum_{a=1}^{m}\frac{\hat{L}_a\rho(t)\hat{L}_a^{\dagger}}{\mathrm{Tr}\left(\sum_{b=1}^{m}\hat{L}_b\rho(t)\hat{L}_b^{\dagger}\right)}-\frac{\eta(t)}{2}\sum_{a=1}^{n}\lbrace \hat{L}_a^{\dagger}\hat{L}_a,\rho(t)\rbrace.
\end{equation}
If we set $m=n$ which means we measure the system on a complete basis,  we will find that the normalization factor $\mathrm{Tr}\left(\sum_{b=1}^{n}\hat{L}_b\rho(t)\hat{L}_b^{\dagger}\right)=1$ since we assume the completeness condition of $\hat{L}_a$ in the Eq.~(\ref{complete}). In this case, the Eq.~(\ref{post_selection_partial}) is the same as the case without post-selection in the Eq.~(\ref{without_post}). 

For the general $m < n$, the normalization factor $\sum_{b=1}^{m}\mathrm{Tr}[\hat{L}_b\rho(t+\delta t)\hat{L}_b^{\dagger}]$ is generally not one and depends on $\rho(t)$, and this is a non-linear differential equation of $\rho(t)$ different from the Lindblad master equation. The non-trivial normalization factor given by the post-selection process leads to a different differential equation of the density matrix. 

\section{The general properties of the generalized Lindblad equation}

A dynamical equation of the density matrix should preserve the Hermitian, unit trace, and positive definiteness properties of the density matrix. In the following, we will show that these three properties of the density matrix are preserved under the evolution of our generalized Lindblad equation. 

We find that only the second term on the left-hand side of the Eq.~(\ref{post_selection_partial}) is different from the original Lindblad master equation, and we know the original Lindblad master equation preserves these three properties. Hence, we can simply compare the difference between the second term on the right-hand side of our generalized Lindblad equation and that of the original Lindblad equation.

Firstly, since $\sum_{a=1}^{m}\frac{\hat{L}_a\rho(t)\hat{L}_a^{\dagger}}{\mathrm{Tr}\left(\sum_{b=1}^{m}\hat{L}_b\rho(t)\hat{L}_b^{\dagger}\right)}$ is Hermitian, and $\mathrm{Tr}\left[\sum_{a=1}^{m}\frac{\hat{L}_a\rho(t)\hat{L}_a^{\dagger}}{\mathrm{Tr}\left(\sum_{b=1}^{m}\hat{L}_b \rho(t)\hat{L}_b^{\dagger}\right)}\right]=\mathrm{Tr}\left[\frac{1}{2}\sum_{a=1}^{n}\lbrace \hat{L}_a^{\dagger}\hat{L}_a,\rho(t)\rbrace\right]$ $=1$, it is easy to prove that $\rho(t)$ is Hermitian and preserves its unit trace along this evolution.

Secondly, the proof of positive definiteness is more involved, and it is as follows. We use the operator-sum representation of quantum channel \cite{quantum_information_note}:
\begin{equation}
	\rho(t+\diff t)=\epsilon_{\diff t}[\rho(t)]=\sum_{a=0}^{n}\hat{M_a}\rho(t)\hat{M}_a^{\dagger}.
\end{equation}
Here, 
\begin{equation}
	\begin{cases} 
		\hat{M}_0&= \hat{I}+(-i\hat{H}+\hat{K})\diff t \\
		\hat{M}_a&= \sqrt{\eta(t)}\hat{L}_a\sqrt{\diff t},     \ \ \ \  a\ne 0
	\end{cases}
\end{equation}
with $\hat{L}_a$ representing quantum jump operators, and $\hat{I}$ representing the identity operator. Here, $\hat{K}=-\frac{1}{2}\eta(t)\sum_{a=1}^n\hat{L}_a^{\dagger}\hat{L}_a$. It is straightforward to see that this operator sum representation of quantum channel is equal to the Lindblad master equation the Eq.~(\ref{without_post}). 

In our case, the density operator can be written in a similar operator-sum representation,
\begin{equation}
	\label{operator_sum_selection}
	\rho(t+\diff t)=\epsilon_{\diff t}^{p}[\rho(t)]=\sum_{b=0}^{m}\hat{M_b}\rho(t)\hat{M}_b^{\dagger}
\end{equation}
with
\begin{equation}
	\label{operator_sum_partial}
	\begin{cases} 
		\hat{M}_0&  =   \hat{I}+(-i\hat{H}+\hat{K})\diff t \\
		\hat{M}_b&  =\sqrt{\alpha_t\eta(t)}\hat{L}_b\sqrt{\diff t},  \ \ \ \         b=1,...,m.
	\end{cases}
\end{equation}
Here, $\hat{K}=-\frac{1}{2}\eta(t)\sum_{a=1}^n\hat{L}_a^{\dagger}\hat{L}_a$, and we also assume the completeness condition the Eq.~(\ref{complete}). Here, the superscript $p$ in the Eq.~(\ref{operator_sum_partial}) denotes the case with post-selection, and $\alpha_t=\left[\mathrm{Tr}\left(\sum_{b=1}^{m}\hat{L}_b\rho(t)\hat{L}_b^{\dagger}\right)\right]^{-1}$ is the normalization factor attributed to post-selection.  Also, since $m\le n$, the case where the system is being projected on the complete basis is included in the Eq.~(\ref{operator_sum_selection}). 
We assume that at time $t$, the density matrix is positive definite, i.e. $\rho(t)$ can be decomposed as $\rho(t)=\sum_j p_j |\psi_j\rangle\langle\psi_j|$, $p_j\ge0$. Thus, we find that
\begin{equation}
\langle \phi|\rho(t+\diff t)|\phi \rangle
		=\sum_{b=0}^n \langle\phi|\hat{M}_b\rho(t)\hat{M}_b^{\dagger}|\phi\rangle =\sum_{b=0}^n\sum_j p_j |\langle\psi_j|\hat{M}_b^{\dagger}|\phi\rangle|^2 \ge0, \  \ \forall  \phi.
\end{equation}
Therefore, we conclude that $\rho(t+\diff t)$ is positive definite provided that $\rho(t)$ is positive definite. This completes the proof.

\section{Entanglement entropy in the MIPT \label{entanglement_section}}
We then consider entropy dynamics in the MIPT. We here simply focus on studying the second R\'enyi entropy defined as: $S^{(2)}=-\log \left[\mathrm{Tr}(\rho^2)\right]$. We first notice that the Eq.~(\ref{post_selection_partial}) is not able to describe entropy dynamics along this process. The reason is as follows. Since the measurement will produce different results, the density matrix of the system during the evolution process is the summation of the density matrix corresponding to the different measurement results, and the weight of each result is its probability of it. We divide the system into subsystems $A$ and $B$, and the reduced density matrix can be written as $\rho_A=\sum_c p_c  \text{Tr}_{B}\rho_c=\sum_c p_c \rho_{A,c}$. Here, $p_c$ is the probability of getting $\rho_{c}$, and it satisfies $\sum_c p_c=1$. Therefore, it can be seen that the entanglement entropy comes from two parts: the entanglement entropy of the system corresponding to different measurement results $-\log\lbrace  \text{Tr} \lbrack\rho_{A,c}^2 \rbrack \rbrace$, and the entropy corresponding to the probability distribution of different measurement results $\{p_c\}$. 
In the MIPT, we are only concerned with the former part of entropy, and therefore we need to exclude the latter part. If the density matrix under the dynamic evolution described by the Eq.~(\ref{post_selection_partial}) is directly used in calculating entropy,  we calculate the entanglement entropy of the mixed state obtained by the average of different measurement results:
\begin{equation}
	\label{s_total}
	S^{\text{total}}_{A}\equiv-\log\left\{\mathrm{Tr}\left[\left(\sum_{c=1}^m p_c \rho_{A,c}\right)^2\right]\right\} .
\end{equation}
Here, $p_c$ is the probality of getting $\rho_{A,c}$, and it satisfies $\sum_c p_c=1$. 
Notice that not only the former part of entropy but also the latter part of entropy is included here. Therefore, the dynamics of entropy cannot be described only by the equation the Eq.~(\ref{post_selection_partial}). 

To exclude the latter part of entropy in MIPT, we define a new type of entanglement entropy $S_A^{\text{new}}$ in this process as
\begin{equation}
	\label{S_A_probability}
	S_A^{\text{new}}=-\log  \left\{ \sum_c \tilde{p_c} \left[ \mathrm{Tr}_A\rho_{A,c}^2 \right] \right\}.
\end{equation}
Here, $\tilde{p_c}= \frac{p_c^2}{ \sum_{c'} p_{c'}^2}$, and it also satisfies $\sum_c \tilde{p_c}=1$. 
To see that this definition of entanglement entropy the Eq.~\eqref{S_A_probability} does not take into account the entropy coming from the probability distribution $\lbrace p_c\rbrace$ in contrast to the Eq.~\eqref{s_total}, we consider a simple example. We assume that the system after evolution has probability $p_1=\frac{1}{2}$ to be in $\rho_{A,1}$ with $\mathrm{Tr}[\rho_{A,1}^2]=1$, and  probability $p_2=\frac{1}{2}$ to be in $\rho_{A,2}$ with $\mathrm{Tr}[\rho_{A,2}^2]=1$. This means that in both cases, there is no entanglement entropy between subsystem $A$ and $B$. Thus, the entropy only comes from the probability distribution $\lbrace p_c \rbrace=\lbrace \frac{1}{2},\frac{1}{2}\rbrace$. Also, we assume $\mathrm{Tr}(\rho_{A,1}\rho_{A,2})<1$ which means $\rho_{A,1}$ and $\rho_{A,1}$ are not the same, then we obtain $	S_A^{\text{new}}=-\log[\frac{1}{2}+\frac{1}{2}]=0$ from the definition in the Eq.~(\ref{S_A_probability}). However, if we use the $S^{\text{total}}_A$ defined in the Eq.~(\ref{s_total}) to calculate the entanglement entropy, we obtain $S_A^{\text{total}}=-\log[\frac{1}{4}+\frac{1}{2}\mathrm{Tr}(\rho_{A,1}\rho_{A,2})+\frac{1}{4}]=-\log[\frac{1}{2}+\frac{1}{2}\mathrm{Tr}(\rho_{A,1}\rho_{A,2})]>0$. Therefore, this definition of entanglement entropy includes the entropy coming from the classical distribution $\lbrace p_c\rbrace$. The proof of a more general case is written in the supplementary material \cite{Supplementary}. 

Moreover, when all density matrix $\rho_{A,c}$ are mutually orthogonal, it is straightforward to prove that $S_A^{\text{total}}\geq 	S_A^{\text{new}}$ where the equality is taken when there is only one outcome with probability $p_1=1$. That means $S_A^{\text{total}}-	S_A^{\text{new}}$ is non-negative as long as the entropy of probability distribution $\lbrace p_c \rbrace$ is non-zero. The details of this proof are written in the supplementary material \cite{Supplementary}.

\section{The application of generalized Lindblad equation on the MIPT}
In this section, we show that the entanglement entropy defined in the Eq.~\eqref{S_A_probability} can be obtained from a density matrix defined on a double space, and this double space density matrix's evolution is governed by a generalized Lindblad equation that we will propose later.  

\begin{figure}[h] 
	\centering 
	\includegraphics[width=0.6\textwidth]{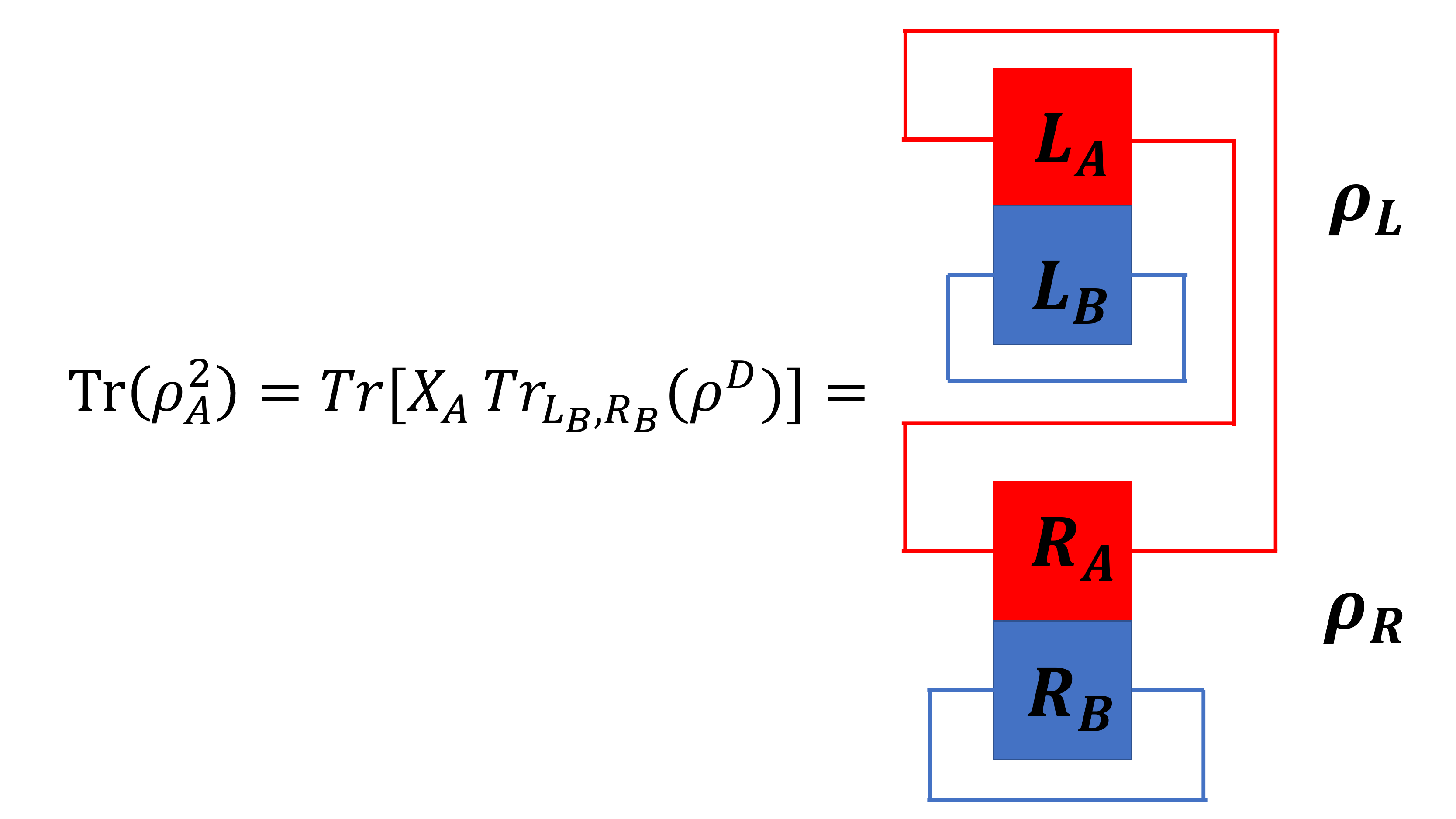} 
	\caption{The schematic diagram of calculating entanglement entropy from the double state density matrix $\rho^D$.} 
	\label{tr_rho_fig} 
\end{figure}

The dynamics of the second R\'enyi entropy can be mapped to the dynamics of a wave function defined on a double space, and therefore the second R\'enyi entropy dynamics is more straightforward when it is written on a double space \cite{Renyi,Double_density}. We denote the two copies of the system on double space as the left(L) and the right(R) system, and we use $\rho^{D}$ to denote the total density matrix of the double system. Given an initial density matrix $\rho=\sum_{mn}\rho_{mn}|m\rangle \langle n|$, the double state density matrix $\rho^D$ is given by:
\begin{equation}
	\rho^D=\rho \otimes \rho=\sum_{mnst}\rho^D_{mn,st}|m\rangle_L\otimes|s\rangle_{R} \langle n|_L \otimes\langle t|_{R}
\end{equation}
with $\rho^D_{mn,st}=\rho_{mn}\rho_{st}$. We can divide the system into subsystems $A$ and $B$, and derive the entanglement entropy of subsystem $A$ from $\rho_D$. Via a standard double space technique, the single system entropy $S_A\equiv -\log \left[\mathrm{Tr}(\rho_A^2)\right]$ can be represented in the double space density matrix. Using $\mathrm{Tr}_A(\rho_A^2)=\mathrm{Tr}_{L_A,R_A}(X_A\rho_A \otimes \rho_A)=\mathrm{Tr}_{L_A,R_A}\left[X_{A} \mathrm{Tr}_{L_B,R_B} (\rho^D)\right]$ , we have 
\begin{equation}
	\label{SA_definition}
	\begin{aligned}
		S_A=&-\log \lbrace \mathrm{Tr}_{L_A,R_A}\left[X_{A} \mathrm{Tr}_{L_B,R_B} (\rho^D)\right]   \rbrace,
	\end{aligned}
\end{equation}
and it is illustrated in the Fig.~(\ref{tr_rho_fig}).

Here, $\rho_A$ is the reduced density matrix of subsystem A  calculated from the total density matrix as $\rho_A=\mathrm{Tr}_B[\rho]$. Here, $\rho$ is the density matrix of the full system. $X$ is the swap operator defined as $X|\alpha\rangle_L |\beta\rangle_R=|\beta\rangle_L |\alpha\rangle_R$. Here, $X_{A}$ is the swap operator that acts on the subspace $A$. $L_A(L_B)$ is the $A$($B$) subsystem of the $L$ system, and $R_A(R_B)$ is the $A$($B$) subsystem of the $R$ system.

When we consider the second R\'enyi entropy, to exclude the entropy coming from the classical distribution of different measurement results, we can enforce the condition that the L and the R system always collapse to the same state after measurement. In the double space, this condition means that the L and the R system are in the Einstein-Podolsky-Rosen (EPR) state. These above considerations motivate the equation of motion of the total density matrix $\rho^{D}$:
\begin{equation}
	\label{general_density}
		\frac{\partial \rho^D}{\partial t}=-i[\hat{H}^D,\rho^{D}(t)]+\eta(t)\sum_{a=1}^{n}\frac{\hat{L}_{a,L} \hat{L}_{a,R}\rho^{D}(t)\hat{L}_{a,L}^{\dagger}\hat{L}_{a,R}^{\dagger}}{\mathrm{Tr}\left(\sum_{b=1}^{n}\hat{L}_{b,L} \hat{L}_{b,R}\rho^{D}(t)\hat{L}_{b,L}^{\dagger}\hat{L}_{b,R}^{\dagger}\right)}
		-\frac{\eta(t)}{2}\sum_{a.b=1}^{n}\lbrace \hat{L}_{a,L}^{\dagger}\hat{L}_{a,L}\hat{L}_{b,R}^{\dagger}\hat{L}_{b,R},\rho^{D}(t)\rbrace.
\end{equation}
Here, $\hat{L}_{a,L}$ and $\hat{L}_{a,R}$ represent the jump operators acting on the $L$ and $R$ system respectively. Also, $\hat{H}^D=\hat{H}\otimes \hat{I} +\hat{I}\otimes\hat{H}$, and $\hat{I}$ denotes the identity operator. This equation can be obtained by replacing $\rho$ by $\rho^D$,$\hat{H}$ by $\hat{H}^D$ and $\hat{L}_a$ by $\hat{L}_{a,L} \hat{L}_{a,R}$ in the Eq.~(\ref{post_selection_partial}). Similar to the discussion in the previous section, the numerator of the second term on the right-hand side $\sum_{a=1}^n \hat{L}_{a,L} \hat{L}_{a,R}\rho^{D}(t)\hat{L}_{a,L}^{\dagger}\hat{L}_{a,R}^{\dagger}$ means that the double system is being projected on the EPR state since the quantum jump operator on the $L$ and $R$ system always being the same. It is a sharp contrast to the case where the numerator is chosen as $\sum_{a,b=1}^n \hat{L}_{a,L} \hat{L}_{b,R}\rho^{D}(t)\hat{L}_{a,L}^{\dagger}\hat{L}_{b,R}^{\dagger}$. Since $\sum_{a,b=1}^n \hat{L}_{a,L} \hat{L}_{b,R}\rho^{D}(t)\hat{L}_{a,L}^{\dagger}\hat{L}_{b,R}^{\dagger}$ describes the system been projected on the complete basis of the double space, whereas EPR states are partial basis of the double space. The denominator on the right hand side $\mathrm{Tr}\left(\sum_{b=1}^{n}\hat{L}_{b,L} \hat{L}_{b,R}\rho^{D}(t)\hat{L}_{b,L}^{\dagger}\hat{L}_{b,R}^{\dagger}\right)$ is the normalization factor resulting from post-selection. It is easy to see that this normalization factor is also non-trivial, making this equation a non-linear differential equation of $\rho^D$.

The generalized Lindblad equation Eq.~\eqref{general_density} together with the entropy calculated through the double space technique the Eq.~\eqref{SA_definition} are the central results of this paper. The essential property of the MIPT can be satisfied with this scheme, and the post-selection is explicitly embedded in the EPR state projection condition in the Eq.~\eqref{general_density}.
These arguments rely on two steps:
\begin{enumerate}
	\item Firstly, we will prove that the entropy formula the Eq.~\eqref{SA_definition} for the evolved density matrix the Eq.~\eqref{general_density} can be rewritten in the form of $S_A^{\text{new}}$ defined in the Eq.~\eqref{S_A_probability}. This argument fundamentally requires the EPR state projection structure achieved by double space and the choice of Lindblad jump operators in the Eq.~\eqref{general_density}.
	\item Secondly, as we find previously, $S_A^{\text{new}}$ in the Eq.~\eqref{S_A_probability} and the $S^{\text{total}}_A$ in the Eq.~\eqref{s_total} are conceptually different when describing the entanglement entropy. $S_A^{\text{new}}$ can exclude the entropy coming from the classical distribution $\lbrace p_c\rbrace$ by excluding the cross term $p_c \rho_c p_{c'} \rho_{c'} $ with $c\neq c'$ when calculating the entropy. In the aforementioned literature about MIPT \cite{percolation,critical_exponent_quantum_trajectory,phase_diagram}, such exclusions are implemented by post-selection, while here it was naturally embedded in the framework of the generalized Lindblad equation.
\end{enumerate}

We will then give a proof of the first argument. Similar to the previous section, the evolution of the double system density matrix can be represented in the operator-sum form
\begin{equation}
	\label{double_operator_sum}
	\rho^D(t+\diff t)=\epsilon_{\diff t}[\rho^D(t)]=\sum_{b=0}^{m}\hat{M}_b^D\rho_0^D(t)\hat{M}_b^{D \dagger}
\end{equation}
with
\begin{equation}
	\label{double_channel_def}
	\begin{cases} 
		\hat{M}_0^D&  =  \hat{I}^D+(-i\hat{H}^D+\hat{K}^D)\diff t \\
		\hat{M}_b^D&  =\sqrt{\alpha_t^D\eta(t)}\hat{L}_{b,L}\hat{L}_{b,R}\sqrt{\diff t},  \ \ \ \         b=1,...,m.
	\end{cases}
\end{equation}
Here, $\hat{K}^D=\hat{K}_L+\hat{K}_R=-\frac{1}{4}\eta(t)\sum_{a=1}^n\hat{L}_{a,L}^{\dagger}\hat{L}_{a,L}-\frac{1}{4}\eta(t)\sum_{a=1}^n\hat{L}_{a,R}^{\dagger}\hat{L}_{a,R}$, and we also assume the completeness condition the Eq.~(\ref{complete}). 
Then, by only keeping the first order of $\diff t$, we find that $\hat{M}_b^D$ can be rewritten as
\begin{equation}
	\label{double_channel_rewrite}
	\begin{cases} 
		\hat{M}_0^D&  = 	\hat{M}_{0,L} \otimes 	\hat{M}_{0,R}\\
		\hat{M}_b^D&  =	\hat{M}_{b,L} \otimes 	\hat{M}_{b,R},  \ \ \       b=1,...,m
	\end{cases}
\end{equation}
with
\begin{equation*}
	\begin{split}
		\hat{M}_0&= \hat{I}+(-i\hat{H}+\hat{K})\diff t \\ \hat{M}_b&=\left[\alpha_t^D\eta(t)\right]^{\frac{1}{4}}\hat{L}_{b} (\diff t)^{\frac{1}{4}}, \\ 
	\end{split}
\end{equation*}
$	\text{and\ } \alpha_t^D=\left[ \mathrm{Tr}\left(\sum_{b=1}^{n}\hat{L}_{b,L} \hat{L}_{b,R}\rho^{D}(t)\hat{L}_{b,L}^{\dagger}\hat{L}_{b,R}^{\dagger}\right)\right]^{-1}$ is the normalization factor attributed to post-selection in the double space.
Then the entanglement entropy calculated from $\rho_D$ can be written as
\begin{equation}
	\label{SA_double}
	\begin{aligned}
		S_A^D =&-\log \left\{ \mathrm{Tr}_{L_A,R_A}\left[X_{A} \mathrm{Tr}_{L_B,R_B} (\sum_{b=0}^{m}\hat{M}_b^D\rho_0^D\hat{M}_b^{D \dagger})\right] \right\} \\
		=&-\log \{ \sum_{b=0}^{m} \mathrm{Tr}_{L_A,R_A}[X_A \mathrm{Tr}_{L_B,R_B} (\hat{M}_{b,L}\rho_0\hat{M}_{b,L}^{ \dagger})\otimes \quad(\hat{M}_{b,R}\rho_0\hat{M}_{b,R}^{ \dagger})]  \}\\
		=&-\log  \left\{ \sum_{b=0}^m \tilde{p_b} \mathrm{Tr}_{L_A,R_A}\left[X_A\rho_{A,b}\otimes \rho_{A,b} \right] \right\}\\
		=&-\log  \left\{ \sum_{b=0}^m \tilde{p_b} \mathrm{Tr}_{A}\left[\rho_{A,b}^2 \right] \right\}
	\end{aligned}
\end{equation}
with $\tilde{p_b} =\left[\mathrm{Tr}\left(\hat{M}_b\rho_0(t)\hat{M}_b^{\dagger}\right)\right]^{2}$ and $\rho_{A,b}=\frac{\mathrm{Tr}_{B}\left[\hat{M}_{b}\rho_0(t)\hat{M}_{b}^{ \dagger}\right]}{\mathrm{Tr}\left(\hat{M}_b\rho_0(t)\hat{M}_b^{\dagger}\right)}$. Also, if we define $\mathrm{Tr}\left(\hat{L}_b\rho_0(t)\hat{L}_b^{\dagger}\right)= p_b$, then we have $\tilde{p_b} =\alpha_t^D\eta(t)\left[\mathrm{Tr}\left(\hat{L}_b\rho_0(t)\hat{L}_b^{\dagger}\right)\right]^{2}=\eta(t)\frac{p_b^2}{\sum_c p_c^2}$  for $b=1,...,n$. When we consider the case $\eta(t)=1$,which means that the probability of the system being measured per unit time is 1, this definition of $\tilde{p_b}$ is consistent with that of  the new type of entanglement entropy $S_A^{\text{new}}$ in the Eq.~\eqref{S_A_probability}. This completes the proof.

\section{Numerical results \label{Numerical_section}}
In this section, we numerically study the second R\'enyi entropy dynamics of a 1D hard-core Bose Hubbard \cite{Bose_hubbard} system under the Eq.~\eqref{general_density} and \eqref{SA_definition} to show that this is consistent with the understanding that there could be an entanglement phase transition.
The Hamiltonian of the hard-core Bose Hubbard system is
\begin{equation}
	\hat{H}=-J\sum_{\langle i,j\rangle}\hat{b}_{i}^{\dagger}\hat{b}_{j}+U\sum_{\langle i,j\rangle}\hat{n}_{i}\hat{n}_{j}.
\end{equation}
Here, $J$ is the strength of the nearest neighbor hopping, and $U$ is the strength of the nearest neighbor interaction. 

The system is driven by the generalized Lindblad equation the Eq.~(\ref{general_density}) in the double system, and we set $\eta(t)=\gamma$ as a time-independent measurement rate. 
Also, we set the projection measurements as
\begin{equation}
	\hat{L}_{i,0}=\frac{1}{\sqrt{L}}(1-\hat{n}_{i}),\  \hat{L}_{i,1}=\frac{1}{\sqrt{L}}\hat{n}_{i}.
\end{equation}
Here $i=1,2,..., N_s$, and $N_s$ is the total number of sites.  Also, we further normalize the projection operators to satisfy the completeness condition of the measurement basis in Eq.~(\ref{complete}).
These projection operators mean that the environment is measuring the particle number on each site by projecting it on one of the particle number basis $(\vert 0\rangle,\vert 1\rangle)$. Here, $\vert 1\rangle$ denotes the site is occupied, and $\vert 0\rangle$ denotes the site is unoccupied.

In our following numerical calculation, we set $J=U=1, N_s=6, N_b=3$. $N_b$ is the total number of the hard-core boson. We denote the left half of the system as subsystem $A$ and the rest of it as subsystem $B$.  We then calculate the entanglement entropy $S_A$ defined in the Eq.~(\ref{SA_definition}). We choose the initial state as a product state in the particle number basis $|000111\rangle$.
\begin{figure}[h] 
	\centering 
	\includegraphics[width=0.6\textwidth]{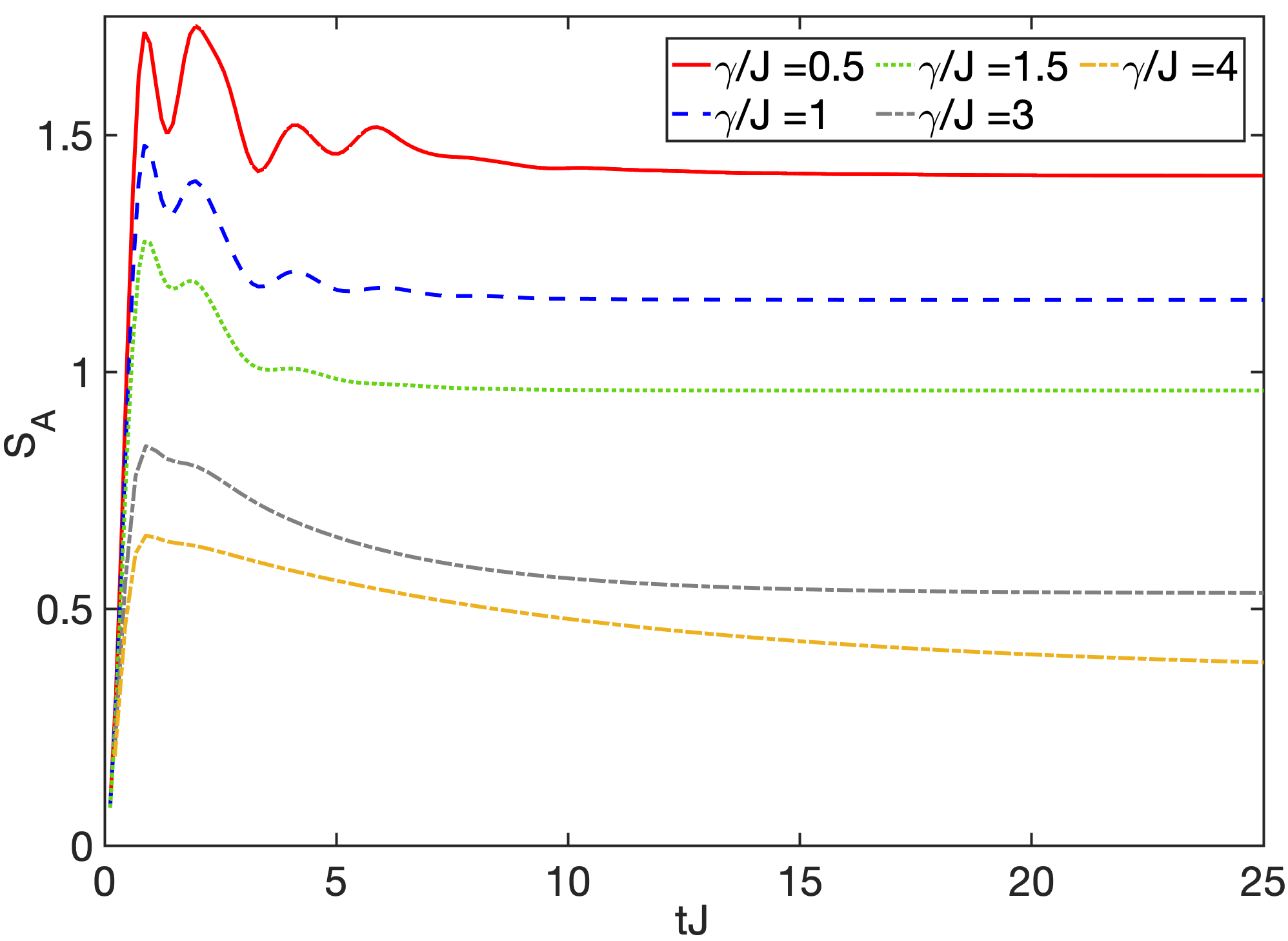} 
	\caption{The dynamics of the entanglement entropy $S_A$ as a
		function of $tJ$. $\gamma$ is the measurement rate. Different curves have different $\gamma$ in the unit of $J$. Here, $U = J$ and the number of sites $N_s = 6$, and the number of bosons $N_b = 3$, and subsystem size $L_A=3$.} 
	\label{time_evolution}
\end{figure}

\begin{figure}[h] 
	\centering 
	\includegraphics[width=0.6\textwidth]{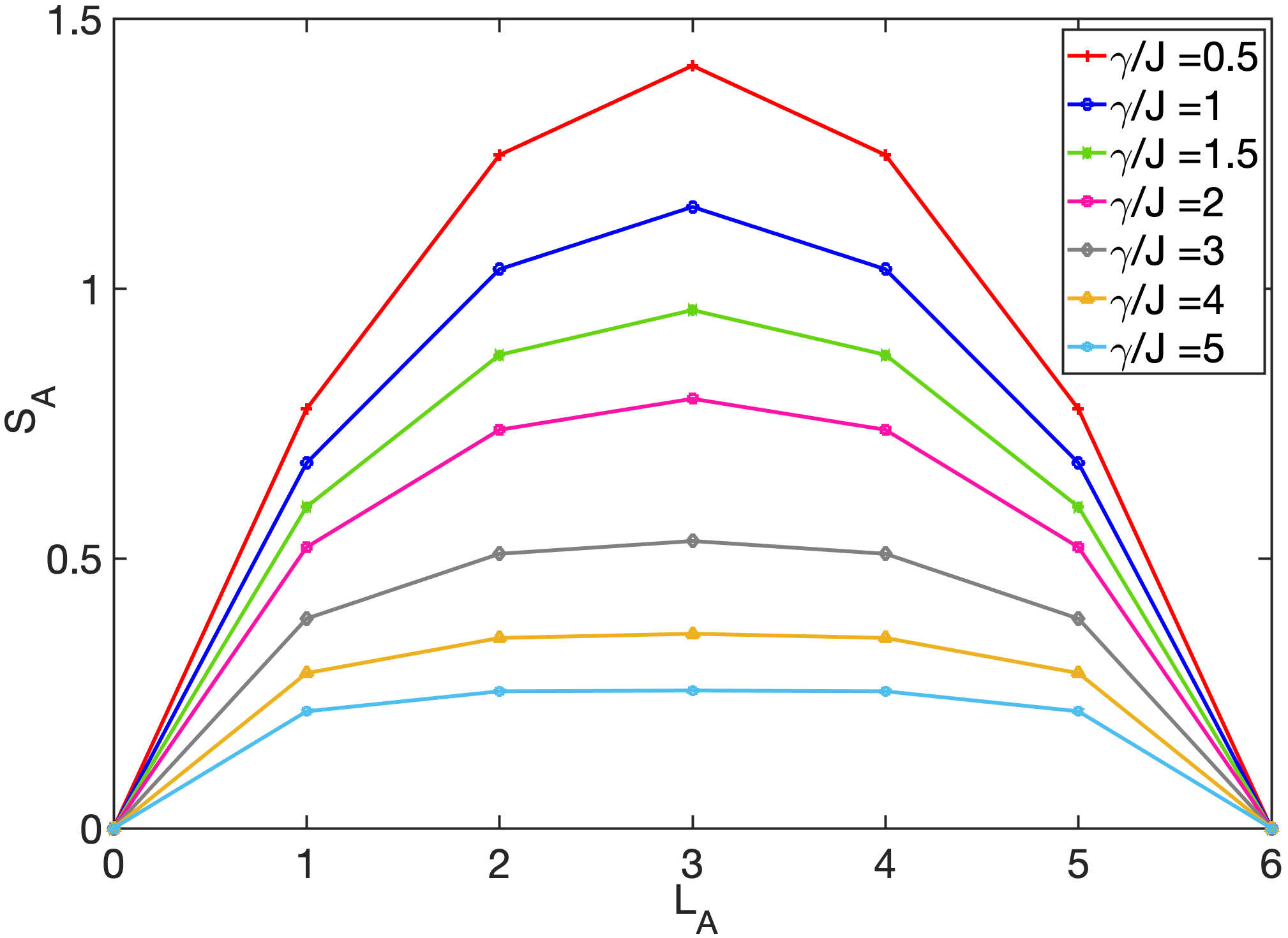} 
	\caption{Saturation value of entanglement entropy $S_A$ of the subsystem $A$ with different subsystem size $L_A$. Different curves have different $\gamma$ in the unit of $J$. Here, $N_s=6$, $N_b=3$, $U/J=1$.} 
	\label{area_law} 
\end{figure} 
As shown in Fig.~\ref{time_evolution}, the entanglement entropy $S_A$ between subsystems $A$ and $B$ first quickly increases as expected in a normal chaotic system. However, it then decreases and saturates to a non-zero value. It indicates that measurements and the following post-selection process decrease the entanglement between subsystems.  Also, we prove in the supplementary material \cite{Supplementary} that the entanglement entropy is the same regardless of whether one computes partial trace over the subsystem $A$ or subsystem $B$, and this explains why entanglement entropy $S_A$ is symmetric to the half system size $L_A$ axis ($L_A=N_s/2=3$) in Fig.~\ref{time_evolution}.

Moreover, from the result in Fig.~\ref{area_law}, we find that when the measurement rate is small ($\gamma/J=0.5$) and $L_A<N_s/2$,  the entanglement entropy between the two subsystems A and B is almost linear in system size $L_A$. Whereas when the measurement rate is large ($\gamma/J=5$) and $L_A<N_s/2$,  the entanglement entropy is almost flat as $L_A$ changes. In the MIPT, as the measurement rate $\gamma$ increases, the entanglement entropy between the two subsystems $A$ and $B$ will change from volume-law to area-law, and our results are consistent with it. As expected, we observe here a reduction of the entanglement entropy as $\gamma/J$ is increased. 

In the section \ref{trajectory_section} of the supplementary material \cite{Supplementary}, we use the quantum trajectory method to numerically calculate the entanglement entropy at different system size $L$, and we find an almost linear growth in the entanglement entropy as the total system size increases at a small value of $\gamma$ ($\gamma/J=0.5$), which indicates that there could be a volume-low scaling at a larger system size. In comparison,  the entanglement entropy varies less as the size of the system increases at large values of $\gamma$ ($\gamma/J=4$), which indicates that there could be a logarithmic critical scaling or an area law (constant) scaling at a larger system size. The details of the quantum trajectory methods that we have used here and more numerical results are also included in the supplementary material \cite{Supplementary}.

Thus, our results are in line with the understanding that there could be an entanglement phase transition in this process. Therefore, we use a concrete example to show that our generalized Lindblad equation the Eq.~(\ref{general_density}) can describe an entanglement phase transition as the measurement rate increases.  Detailed information about the numerical integration technique used in producing Fig.~\ref{time_evolution} and Fig.~\ref{area_law} is added in the supplementary material \cite{Supplementary}.

Also,  the numerical results of entanglement entropy calculated from the single-copy master equation the Eq.~\eqref{post_selection_partial} are added in the supplementary material \cite{Supplementary} for comparison. We find that there is no entanglement phase transition in this process, and there is no surprise since we mentioned in the section \ref{entanglement_section} that the entanglement entropy calculated from the single-copy master equation is $S^{\text{total}}_{A}$ defined in the Eq.~\eqref{s_total}. Since $S^{\text{total}}_{A}$ also includes the entropy corresponding to the probability distribution of different measurement results, there is no entanglement phase transition regarding this entanglement entropy.

\section{Discussions}

In this paper, we derive a generalized Lindblad equation for describing the dynamics of the system under the measurement process followed by post-selection. We emphasize that post-selection is essential in the non-linear differential the Eq.~\eqref{general_density}. Also, the generalized Lindblad equation the Eq.~\eqref{general_density} preserves the Hermitian, unit trace, and positive definiteness of the density matrix. Furthermore, we generalize it to describe the second R\'enyi entropy dynamics in the MIPT, and we use a concrete model to demonstrate that our equation can indeed describe this entanglement transition as the measurement rate increases. Also, our generalized Lindblad equation can be simulated by the quantum trajectory methods \cite{quantum_trajectory1992,quantum_trajectory1993,quantum_trajectory1998,quantum_trajectory2014,quantum_trajectory_PRA1992,quantum_trajectory_book1993,quantum_trajectory_diffusion1992} as the original Lindblad equation, and we consign the details to the Supplemental Material \cite{Supplementary}. Therefore, the numerical simulation of these dissipative dynamics is feasible. 

Moreover, the post-selection provides information about the probability distribution of measurement results. By using the information of the measurement results, we project the system to some pure states. Hence, we decrease the entropy of the system. Moreover, the Holevo information \cite{Holevo} defined as $\chi := S_v(\rho)-\sum_i p_i S_v(\rho_i)$ actually measures how much entropy on average is reduced once we learn the distribution $\{p_i\}$. Here, $S_v$ is the von Neumann entropy. Thus, if we measure von Neumann entropy in the MIPT, we find that the amount of entropy decreased by measurement and post-selection is just the Holevo information. The decrease of entropy by the measurement and post-selection process is owing to gaining accessible information about the measurement results.

The results we have presented here suggest some further directions that are worth exploring. It will be enlightening to analytically calculate the entropy from our generalized Lindblad equation. Also, It will be
interesting to explore the Holevo information in the MIPT and try to understand this phase transition from the perspective of getting accessible classical information. It is also interesting to experimentally realize this generalized Lindblad equation by coupling the system to a bath and designing the form of interaction to satisfy the normalization factor of our generalized Lindblad equation. Therefore, we may further find some direct experimental access to the entanglement phase transition.

\section*{Acknowledgments}

We thank Hui Zhai, Pengfei Zhang, and Tian-Gang Zhou for helpful discussion and for carefully reading the manuscript.

\bibliographystyle{unsrt}

\begin{thebibliography}{50}%
	
	\bibitem{entanglementNahum2019} B. Skinner, J. Ruhman, and A. Nahum, Measurement-Induced Phase Transitions in the Dynamics of Entanglement, Phys. Rev. X 9, 031009 (2019).
	
	\bibitem{entanglementFisher2018} Y. Li, X. Chen, and M. P. A. Fisher, Quantum Zeno Effect and the Many-Body Entanglement Transition, Phys. Rev. B 98, 205136 (2018).
	
	\bibitem{entanglementFisher2019}Y. Li, X. Chen, and M. P. A. Fisher, Measurement-Driven Entanglement Transition in Hybrid Quantum Circuits, Phys. Rev. B 100, 134306 (2019).
	
	\bibitem{entanglementAmosChan2019}A. Chan, R. M. Nandkishore, M. Pretko, and G. Smith, Unitary-Projective Entanglement Dynamics, Phys. Rev. B 99, 224307 (2019).
	
	\bibitem{ultracold_experiment}G. Mazzucchi, W. Kozlowski, S. F. Caballero-Benitez, T. J. Elliott, and I. B. Mekhov, Quantum Measurement-Induced Dynamics of Many-Body Ultracold Bosonic and Fermionic Systems in Optical Lattices, Phys. Rev. A 93, 023632 (2016).
	
	\bibitem{entanglement_weak}M. Szyniszewski, A. Romito, and H. Schomerus, Entanglement Transition from Variable-Strength Weak Measurements, Phys. Rev. B 100, 064204 (2019).
	
	\bibitem{quantum_measure} M. J. Gullans and D. A. Huse, Scalable Probes of Measurement-Induced Criticality, Phys. Rev. Lett. 125, 070606 (2020).
	
	\bibitem{percolation} C.-M. Jian, Y.-Z. You, R. Vasseur, and A. W. W. Ludwig, Measurement-Induced Criticality in Random Quantum Circuits, Phys. Rev. B 101, 104302 (2020).
	
	\bibitem{critical_exponent_Random_circuit} A. Lavasani, Y. Alavirad, and M. Barkeshli, Measurement-Induced Topological Entanglement Transitions in Symmetric Random Quantum Circuits, Nat. Phys. 17, 3 (2021).
	
	\bibitem{critical_exponent_quantum_trajectory} Y. Fuji and Y. Ashida, Measurement-Induced Quantum Criticality under Continuous Monitoring, Phys. Rev. B 102, 054302 (2020).
	
	\bibitem{critical_exponent_Fermion} O. Alberton, M. Buchhold, and S. Diehl, Entanglement Transition in a Monitored Free-Fermion Chain: From Extended Criticality to Area Law, Phys. Rev. Lett. 126, 170602 (2021).
	
	\bibitem{critical_exponent_Fermion2}X. Turkeshi, A. Biella, R. Fazio, M. Dalmonte, and M. Schiró, Measurement-Induced Entanglement Transitions in the Quantum Ising Chain: From Infinite to Zero Clicks, Phys. Rev. B 103, 224210 (2021).
	
	\bibitem{critical_random circuit} A. Zabalo, M. J. Gullans, J. H. Wilson, S. Gopalakrishnan, D. A. Huse, and J. H. Pixley, Critical Properties of the Measurement-Induced Transition in Random Quantum Circuits, Phys. Rev. B 101, 060301 (2020).
	
	\bibitem{critical_circuit_3Dpercolation} X. Turkeshi, R. Fazio, and M. Dalmonte, Measurement-Induced Criticality in $(2+1)$-Dimensional Hybrid Quantum Circuits, Phys. Rev. B 102, 014315 (2020).
	
	
	\bibitem{phase_diagram} S. Sang and T. H. Hsieh, Measurement-Protected Quantum Phases, Phys. Rev. Research 3, 023200 (2021).
	
	\bibitem{statisticRUC} T. Zhou and A. Nahum, Emergent Statistical Mechanics of Entanglement in Random Unitary Circuits, Phys. Rev. B 99, 174205 (2019).
	
	\bibitem{purification} M. J. Gullans and D. A. Huse, Dynamical Purification Phase Transition Induced by Quantum Measurements, Phys. Rev. X 10, 041020 (2020).
	
	\bibitem{quantum_error} S. Choi, Y. Bao, X.-L. Qi, and E. Altman, Quantum Error Correction in Scrambling Dynamics and Measurement-Induced Phase Transition, Phys. Rev. Lett. 125, 030505 (2020).
	
	\bibitem{quantum_error2021}R. Fan, S. Vijay, A. Vishwanath, and Y.-Z. You, Self-Organized Error Correction in Random Unitary Circuits with Measurement, Phys. Rev. B 103, 174309 (2021).
	
	\bibitem{syk_symmetry} S.-K. Jian, C. Liu, X. Chen, B. Swingle, and P. Zhang, Measurement-Induced Phase Transition in the Monitored Sachdev-Ye-Kitaev Model, Phys. Rev. Lett. 127, 140601 (2021).
	
	\bibitem{RTN} R. Vasseur, A. C. Potter, Y.-Z. You, and A. W. W. Ludwig, Entanglement Transitions from Holographic Random Tensor Networks, Phys. Rev. B 100, 134203 (2019).
	
	\bibitem{tensor_network}Z.-C. Yang, Y. Li, M. P. A. Fisher, and X. Chen, Entanglement Phase Transitions in Random Stabilizer Tensor Networks, ArXiv:2107.12376 [Cond-Mat, Physics:Hep-Th, Physics:Quant-Ph] (2021).
	
	\bibitem{all_to_all_circuit}A. Nahum, S. Roy, B. Skinner, and J. Ruhman, Measurement and Entanglement Phase Transitions in All-to-All Quantum Circuits, on Quantum Trees, and in Landau-Ginsburg Theory, PRX Quantum 2, 010352 (2021).
	
	
	\bibitem{MPS} Q. Tang and W. Zhu, Measurement-Induced Phase Transition: A Case Study in the Nonintegrable Model by Density-Matrix Renormalization Group Calculations, Phys. Rev. Research 2, 013022 (2020).
	
	
	\bibitem{RUC} Y. Bao, S. Choi, and E. Altman, Theory of the Phase Transition in Random Unitary Circuits with Measurements, Phys. Rev. B 101, 104301 (2020).
	
	\bibitem{measurement_book}H. M. Wiseman and G. J. Milburn, Quantum Measurement and Control (Cambridge University Press, Cambridge, 2009).
	
	\bibitem{master_equation}J. D. Cresser, S. M. Barnett, J. Jeffers, and D. T. Pegg, Measurement Master Equation, Optics Communications 264, 352 (2006).
	
	\bibitem{measurement_add}C. Presilla, R. Onofrio, and U. Tambini, Measurement Quantum Mechanics and Experiments on Quantum Zeno Effect, Annals of Physics 248, 95 (1996).
	
	\bibitem{open_system} H.-P. Breuer and F. Petruccione, The Theory of Open Quantum Systems (Oxford University Press, Oxford, 2007).
	
		\bibitem{weak_measurement_1988} Y. Aharonov, D. Z. Albert, and L. Vaidman, How the Result of a Measurement of a Component of the Spin of a Spin-1/2 Particle Can Turn out to Be 100, Phys. Rev. Lett. 60, 1351 (1988).
	
	\bibitem{weak_measurement_2021}  A. Biella and M. Schiró, Many-Body Quantum Zeno Effect and Measurement-Induced Subradiance Transition, Quantum 5, 528 (2021).
	
	\bibitem{quantum_Zeno_2020}  K. Snizhko, P. Kumar, and A. Romito, Quantum Zeno Effect Appears in Stages, Phys. Rev. Research 2, 033512 (2020).
	
	\bibitem{quantum_information_note} J. Preskill, Lecture Notes for Physics 229: Quantum Information and Computation, 321 (n.d.).
	
	\bibitem{Renyi} Y.-N. Zhou, L. Mao, and H. Zhai, Renyi Entropy Dynamics and Lindblad Spectrum for Open Quantum Systems, Phys. Rev. Research 3, 043060 (2021).
	
	\bibitem{Double_density} M. Buchhold, Y. Minoguchi, A. Altland, and S. Diehl, Effective Theory for the Measurement-Induced Phase Transition of Dirac Fermions, Phys. Rev. X 11, 041004 (2021).
	
	
	
	
	\bibitem{Bose_hubbard} M. P. A. Fisher, P. B. Weichman, G. Grinstein, and D. S. Fisher, Boson Localization and the Superfluid-Insulator Transition, Phys. Rev. B 40, 546 (1989).
	
	\bibitem{Holevo} A. S. Holevo, “Bounds for the Quantity of Information Transmitted by a Quantum Communication Channel”, Probl. Peredachi Inf., 9:3 (1973), 3-11; Problems Inform. Transmission, 9:3 (1973), 177-183
	
	\bibitem{quantum_trajectory1992}J. Dalibard, Y. Castin, and K. Mølmer, Wave-Function Approach to Dissipative Processes in Quantum Optics, Phys. Rev. Lett. 68, 580 (1992).
	
	\bibitem{quantum_trajectory_PRA1992} R. Dum, P. Zoller, and H. Ritsch, Monte Carlo Simulation of the Atomic Master Equation for Spontaneous Emission, Phys. Rev. A 45, 4879 (1992).
	
	\bibitem{quantum_trajectory_diffusion1992} N. Gisin and I. C. Percival, The Quantum-State Diffusion Model Applied to Open Systems, J. Phys. A: Math. Gen. 25, 5677 (1992).
	
	
	\bibitem{quantum_trajectory_book1993} H. Carmichael, An Open Systems Approach to Quantum Optics, Lecture Notes in Physics Monographs 18, (1993).
	
	
	\bibitem{quantum_trajectory1993} K. Mølmer, Y. Castin, and J. Dalibard, Monte Carlo Wave-Function Method in Quantum Optics, J. Opt. Soc. Am. B, JOSAB 10, 524 (1993).
	
	\bibitem{quantum_trajectory1998} M. B. Plenio and P. L. Knight, The Quantum-Jump Approach to Dissipative Dynamics in Quantum Optics, Rev. Mod. Phys. 70, 101 (1998).
	
	\bibitem{quantum_trajectory2014}A. J. Daley, Quantum Trajectories and Open Many-Body Quantum Systems, Advances in Physics 63, 77 (2014).
	
	\bibitem{Supplementary} See supplementary material for the details of the definition for the double space density matrix. Menawhile, three different definitions of entanglement entropy and their comparasions in the MIPT are also provided. Also, the quantum trajectories method regarding to our generalized Lindblad equation are discussed in it.
	
\end{thebibliography}

\onecolumn
\appendix
\vspace{20pt}

\section{The definition of the double space density matrix from an initial single space density matrix}

\begin{figure}[h]
	\centering 
	\includegraphics[width=0.6\textwidth]{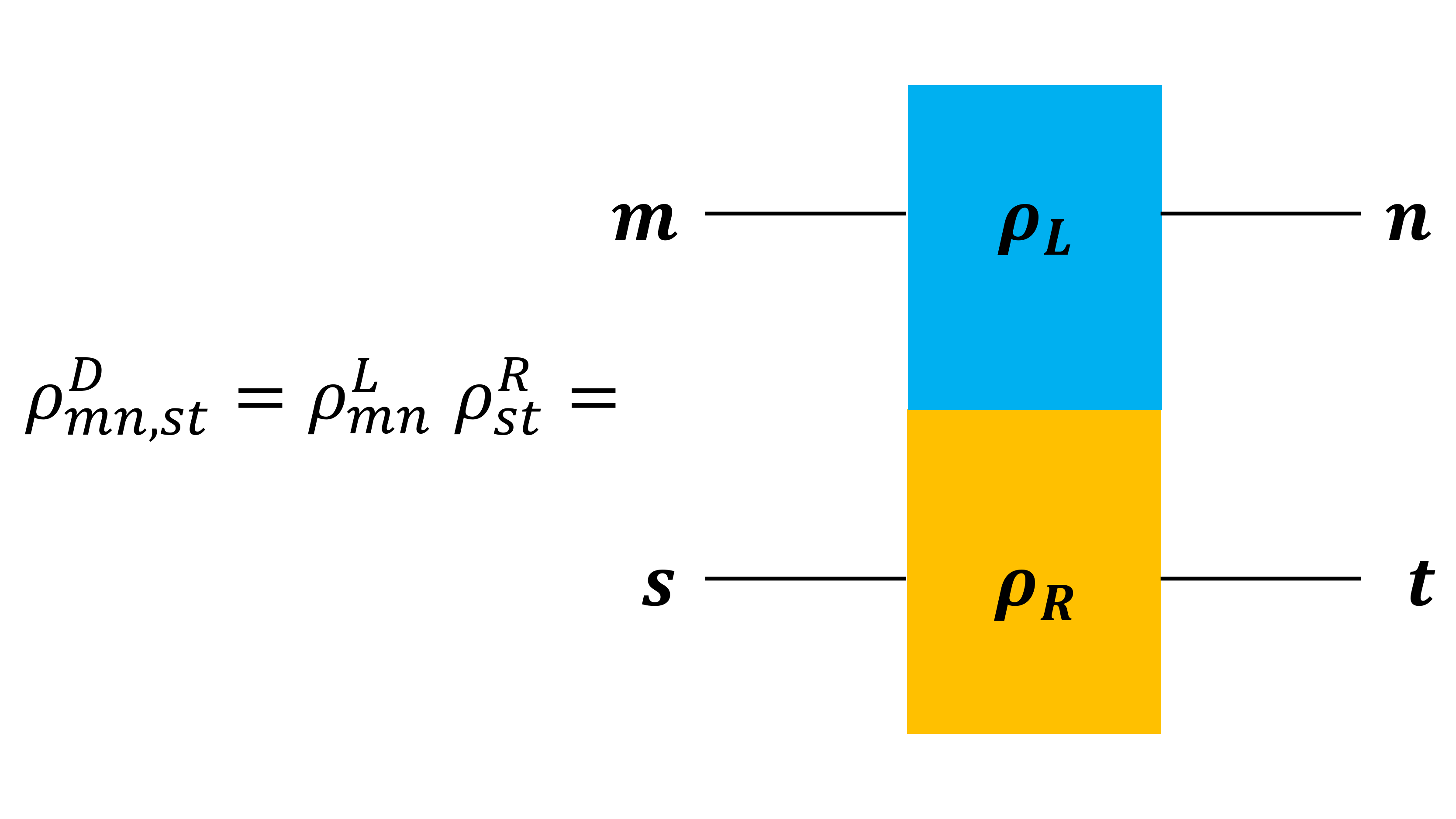} 
	\caption{The schematic diagram of the definition of $\rho^D$.} 
	\label{rhoD_fig} 
\end{figure}

Given an initial density matrix $\rho=\sum_{mn}\rho_{mn}|m\rangle \langle n|$, the double space  density matrix $\rho^D$ is given by:
\begin{equation}
	\rho^D=\sum_{mnst}\rho^D_{mn,st}|m\rangle_L\otimes|s\rangle_{R \ L}\langle n|\otimes_{ R}\langle t|
\end{equation}
with $\rho^D_{mn,st}=\rho_{mn}\rho_{st}$. Here, we assume that the initial double space density matrix is the direct product of two same single space density matrices. $\rho^D$ is illustrated in the Fig.~(\ref{rhoD_fig}).

\section{Three different definitions of entanglement entropy in the measurement-induced phase transition}

In the evolution process together with measurements, measurements will produce different results. Thus,  the density matrix in this process is the summation of the density matrix corresponding to the different measurement results, and each weight of the sum is the probability of that result. Therefore, we can write the density matrix as $\rho=\sum_c p_c \rho_c$.  Here, $c$ represents the different cases of evolution due to different measurement results. If we calculate the entanglement entropy directly from this mixed state density matrix, we obtain  
\begin{equation}
	\label{s_total_suppli}
	S^{\text{total}}=-\log \left\{\mathrm{Tr}\left[(\sum_c p_c \rho_c)^2\right]\right\} .
\end{equation}
Also, we propose a new definition of the second-order R\'enyi entropy in our paper:
\begin{equation}
	\label{s_new}
	S^{\text{new}}=-\log \left\{ \sum_c \tilde{p_c} \mathrm{Tr} \lbrack\rho_{c}^2\rbrack\right\}.
\end{equation}
Here, $\tilde{p_c}= \frac{p_c^2}{ \sum_{c'} p_{c'}^2}$.
There is also another kind of the second-order R\'enyi entropy that people have used in the MIPT \cite{percolation,critical_exponent_quantum_trajectory}:
\begin{equation}
	\label{s_old}
	S^{\text{old}}=-\sum_c p_c \log \lbrace \mathrm{Tr} \lbrack\rho_{c}^2\rbrack\rbrace.
\end{equation}

\section{The comparison of three different definitions of entanglement entropy in the MIPT}

The entropy in MIPT comes from two parts: the entropy of the system corresponding to different measurement results $-\log\lbrace  \text{Tr} \lbrack\rho_c^2 \rbrack \rbrace$, and the entropy of the probability distribution of different measurement results $\{p_c\}$. In the MIPT, we only care about the former part of entropy, and therefore we need to exclude the latter part. 

We will show that both the $S^{\text{new}}$ and $S^{\text{old}}$ can exclude the latter part of entropy, whereas $S^{\text{total}}$ can not. For instance, we assume that the system after evolution has a probability $p_c$ to be in the case $c$, and the density matrix in the case $c$ is $\rho_c$. Here, $\forall c, p_c>0$, and $\sum_{c=1}^n p_c=1 \text{ with } n>1$. We assume that every $\rho_c$ satisfies $\mathrm{Tr}[\rho_c^2]=1$, and this means that in each case, $\rho_c$ is a pure state density matrix. Meanwhile, we assume $\mathrm{Tr}(\rho_{c}\rho_{c'})<1$ for $ c\neq c'$, and this means $\rho_{c}$ and $\rho_{c'}$ are not the same. Thus, the entropy only comes from the probability distribution $\lbrace p_c \rbrace=\lbrace p_1,p_2,...,p_n\rbrace$. From the definition in the Eq.~\eqref{s_total_suppli}, we have
\begin{equation}
	\begin{aligned}
		S^{\text{total}}=&-\log\lbrace \mathrm{Tr}[(\sum_{c=1}^n p_c \rho_c)^2]\rbrace\\
		=&-\log\left[  \sum_{c=1}^n p_c^2 \mathrm{Tr}(\rho_c^2)+2\sum_{c < c'}p_c p_{c'}\mathrm{Tr}(\rho_c \rho_{c'})  \right] \\
		>&-\log\left[ \sum_{c=1}^n p_c^2 \mathrm{Tr}(\rho_c^2)+2\sum_{c< c'}p_c p_{c'}   \right] \\
		=&-\log\left[  \sum_{c=1}^n p_c^2+2\sum_{c< c'}p_c p_{c'} \right]\\
		=&-\log\left[  (\sum_{c=1}^n p_c )^2  \right]=0.
	\end{aligned}
\end{equation}
Here, we use $\mathrm{Tr}(\rho_{A,c}\rho_{A,c'})<1$ for $ c\neq c'$ to get the first inequality. Thus, we obtain $S^{\text{total}}>0$.  

In comparison with $S^{\text{total}}$, from the definition in the Eq.~\eqref{s_new}, we have $S^{\text{new}}=-\log \lbrack\sum_i \tilde{p_i} \mathrm{Tr} (1)\rbrack=0$. Also, from the definition in the Eq.~\eqref{s_old}, we have $S^{\text{old}}=-\sum_i p_i \log \lbrack \mathrm{Tr} (1)\rbrack=0$. Therefore, we find that $S^{\text{total}}$ takes into account the entropy of probability distribution $\lbrace p_c \rbrace=\lbrace p_1,p_2,...,p_n\rbrace$, whereas $S^{\text{new}}$ and $S^{\text{old}}$ do not.

Also, we find that if the density matrix in each case has the same purity and occurs with the same probability, then $S^{\text{new}}$ and $S^{\text{old}}$ are the same. In this case, we have $p_c=\frac{1}{n},\mathrm{Tr}[\rho_c^2]=a$ for $c=1,...,n$. Then, we have $S^{\text{new}}=-\log (\sum_{c=1}^n\frac{1}{n}a)=-\log(a)$, and $S^{\text{old}}=-\sum_{c=1}^n \frac{1}{n} \log a=-\log\lbrack a^{(n\times\frac{1}{n})}\rbrack=-\log(a)$. Thus, we find that  $S^{\text{new}}$ and $S^{\text{old}}$ are the same in this case.

\section{The proof of an inequality between $S^{\text{total}}$ and $S^{\text{new}}$}

In this section, we will prove that when all density matrix $\rho_{c}$ are mutually orthogonal, we have $S^{\text{total}}\geq S^{\text{new}}$ where the equality is taken when there is only one outcome with probability $p_1=1$.  Start from the definition $	S^{\text{total}}$, we have
\begin{equation}
	\begin{aligned}
		S^{\text{total}}=&-\log\lbrace \mathrm{Tr}[(\sum_{c=1}^n p_c \rho_c)^2]\rbrace\\
		=&-\log\left[  \sum_{c,c'=1}^n p_c p_{c'} \mathrm{Tr}(\rho_c\rho_{c'})  \right] \\
		=&-\log\left[  \sum_{c}^n p_c^2  \mathrm{Tr}(\rho_c^2)  \right] \\
		\geq&-\log\left[  \sum_{c}^n \tilde{p_c}  \mathrm{Tr}(\rho_c^2)  \right] \\
		=&S^{\text{new}}.
	\end{aligned}
\end{equation}
Here, we use the $\mathrm{Tr}(\rho_c\rho_{c'})=0$ for $c\neq c'$ to obtain the third equality. Here, Also, we use $\tilde{p_c}= \frac{p_c^2}{ \sum_{c'} p_{c'}^2}\geq p_c^2$ to obtain the fourth line. The equality in the fourth line is taken when there is only one outcome with probability $p_1=1$. Thus, we complete the proof.

\section{The quantum trajectories method}
	\label{trajectory_section}
In this section, we introduce the quantum trajectories method regarding our generalized Lindblad equation the Eq.~\eqref{general_density}. The quantum trajectories method involves rewriting the master equation as a stochastic average over individual trajectories. It is an efficient tool for numerically simulating dissipative dynamics. Similar to the original Lindblad Master equation, our generalized Lindblad equation  can be expressed in an alternative form:
\begin{equation}
		\label{Lindblad_alternative}
		\frac{\partial \rho^D}{\partial t}=-i\left(\hat{H}^D_{\text{eff}}\rho(t)-\rho(t)\hat{H}^{D \dagger}_{\text{eff}}\right)+\eta(t)\sum_{a=1}^{n}\frac{\hat{L}_{a,L} \hat{L}_{a,R}\rho^{D}(t)\hat{L}_{a,L}^{\dagger}\hat{L}_{a,R}^{\dagger}}{\mathrm{Tr}\left(\sum_{b=1}^{n}\hat{L}_{b,L} \hat{L}_{b,R}\rho^{D}(t)\hat{L}_{b,L}^{\dagger}\hat{L}_{b,R}^{\dagger}\right)}.
\end{equation}
Here $\hat{H}^D_{\text{eff}}$ is the effective Hamiltonian defined as
\begin{equation}
	\label{H_eff}
	\begin{aligned}
		\hat{H}^D_{\text{eff}}=\hat{H}\otimes \hat{I} +\hat{I}\otimes\hat{H}-i\frac{\eta(t)}{2}\sum_{a,b}  \hat{L}_{a,L}^{\dagger}\hat{L}_{a,L}\hat{L}_{b,R}^{\dagger}\hat{L}_{b,R}.
	\end{aligned}
\end{equation}
First, we start from a double system initial state $|\phi^D(t)\rangle$, and compute its evolution under the effective Hamiltonian after a small time-step $\delta t$:
\begin{equation}
	|\phi^{(1),D}(t+\delta t)\rangle=(1-iH_{\text{eff}}^D\delta t)|\phi^D(t)\rangle.
\end{equation}
Then, we compute the norm of this wave function at time $t+\delta t$.
\begin{equation}
	\langle\phi^{(1),D}(t+\delta t)	|\phi^{(1),D}(t+\delta t)\rangle\equiv1-\delta p
\end{equation}
where 
\begin{equation}
	\delta p=\eta(t)	\langle\phi^{D}(t)	|i(H_{\text{eff}}^D-H_{\text{eff}}^{D \dagger})|\phi^{D}(t)\rangle\delta t
	=\eta(t)	\langle\phi^{D}(t)	|\sum_{a,b}  \hat{L}_{a,L}^{\dagger}\hat{L}_{a,L}\hat{L}_{b,R}^{\dagger}\hat{L}_{b,R}|\phi^{D}(t)\rangle\delta t.
\end{equation}
Here, we assume the completeness condition of Eq.~(\ref{complete}). Thus, we further have
\begin{equation}
	\delta p=\eta(t)	\delta t.
\end{equation}
Second, we choose the propagated state stochastically in the following manner:
\\1. With probability $1-\delta p$, the wave function at $t+\delta t$ is chosen as the one that  evolves under the effective Hamiltonian with a normalization factor accordingly:
\begin{equation}
	|\phi^{D}(t+\delta t)\rangle=\frac{|\phi^{(1),D}(t+\delta t)\rangle}{\sqrt{1-\delta p}}.
\end{equation}
2. With probability $\delta p$,  the wave function at $t+\delta t$ is chosen as the one that jumps to some particular quantum channel $a$:
\begin{equation}
	|\phi^{D}(t+\delta t)\rangle=\frac{\hat{L}_{a,L}\hat{L}_{a,R}|\phi^{D}( t)\rangle}{\sqrt{\frac{\delta p_a}{\delta t}}}
\end{equation}
where $\delta p_a=\delta t\langle \phi^{D}( t)|\hat{L}_{a,R}^{\dagger}\hat{L}_{a,L}^{\dagger}\hat{L}_{a,L}\hat{L}_{a,R}|\phi^{D}(t)\rangle$. Here, each quantum channel $a$ is chosen with a probability
$\Pi_a = \frac{\delta p_a}{\sum_b \delta p_b}\delta p$.

\begin{figure}[h] 
	\centering 
	\includegraphics[width=0.6\textwidth]{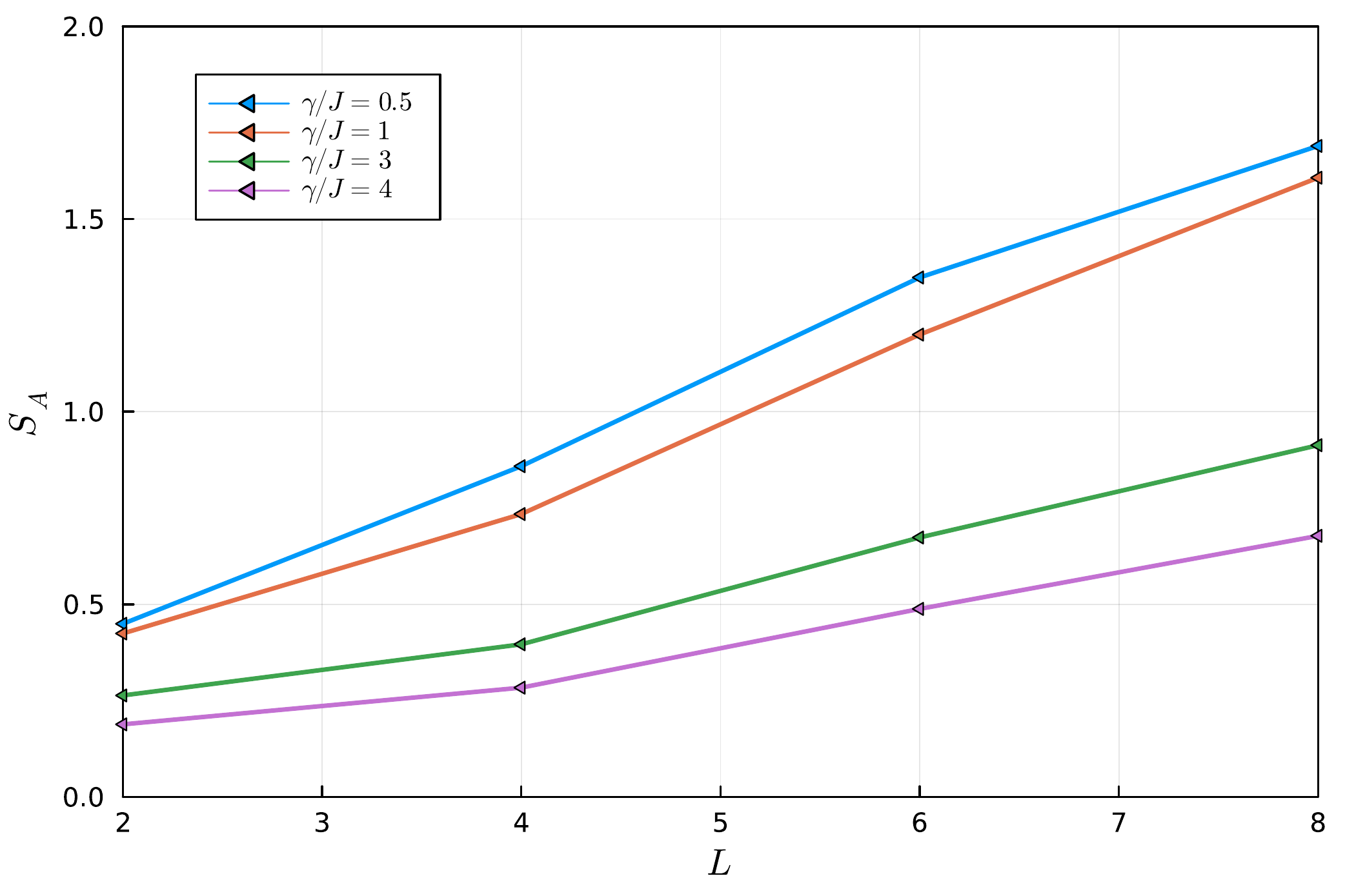} 
	\caption{The longtime saturation value of entanglement entropy $S_A$ of the subsystem $A$ with different system size $N_s$ and fixing $L_A = L/4$, total number of bosons $N_b=L/2$.  Different curves have different $\gamma$ in the unit of $J$. Here, $U/J=1$. We choose a set of $N =100$ quantum trajectories for $L=8$ and $N =1000$ quantum trajectories for $L=2,4,6$.We choose the initial state as a product state in the particle number basis. We choose the initial state $|01\rangle$ for L=2, $|0011\rangle$ for L=4, $|000111\rangle$ for L=6 and $|00001111\rangle$ for L=8.} 
	\label{saturate_trajectory}
\end{figure}

\begin{figure}[h] 
	\centering 
	\includegraphics[width=0.6\textwidth]{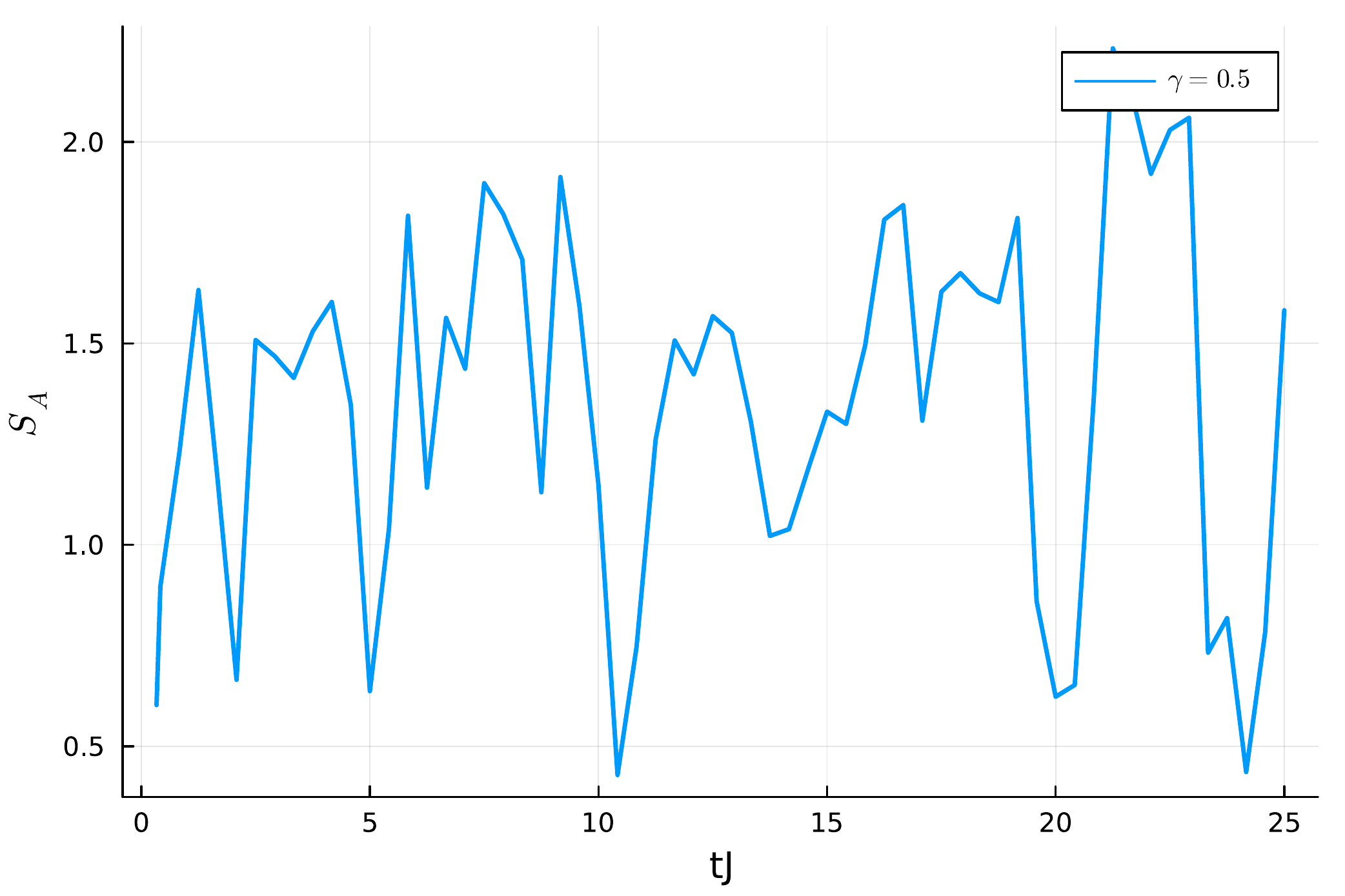} 
	\caption{The entanglement entropy computed on the single realization of a quantum trajectory. Here, $U = J$ and the number of sites $N_s = 6$, and the number of bosons $N_b = 3$, and subsystem size $L_A=3$. The measurement rate is $\gamma/J =0.5$.} 
	\label{single_1}
\end{figure}

\begin{figure}[h] 
	\centering 
	\includegraphics[width=0.6\textwidth]{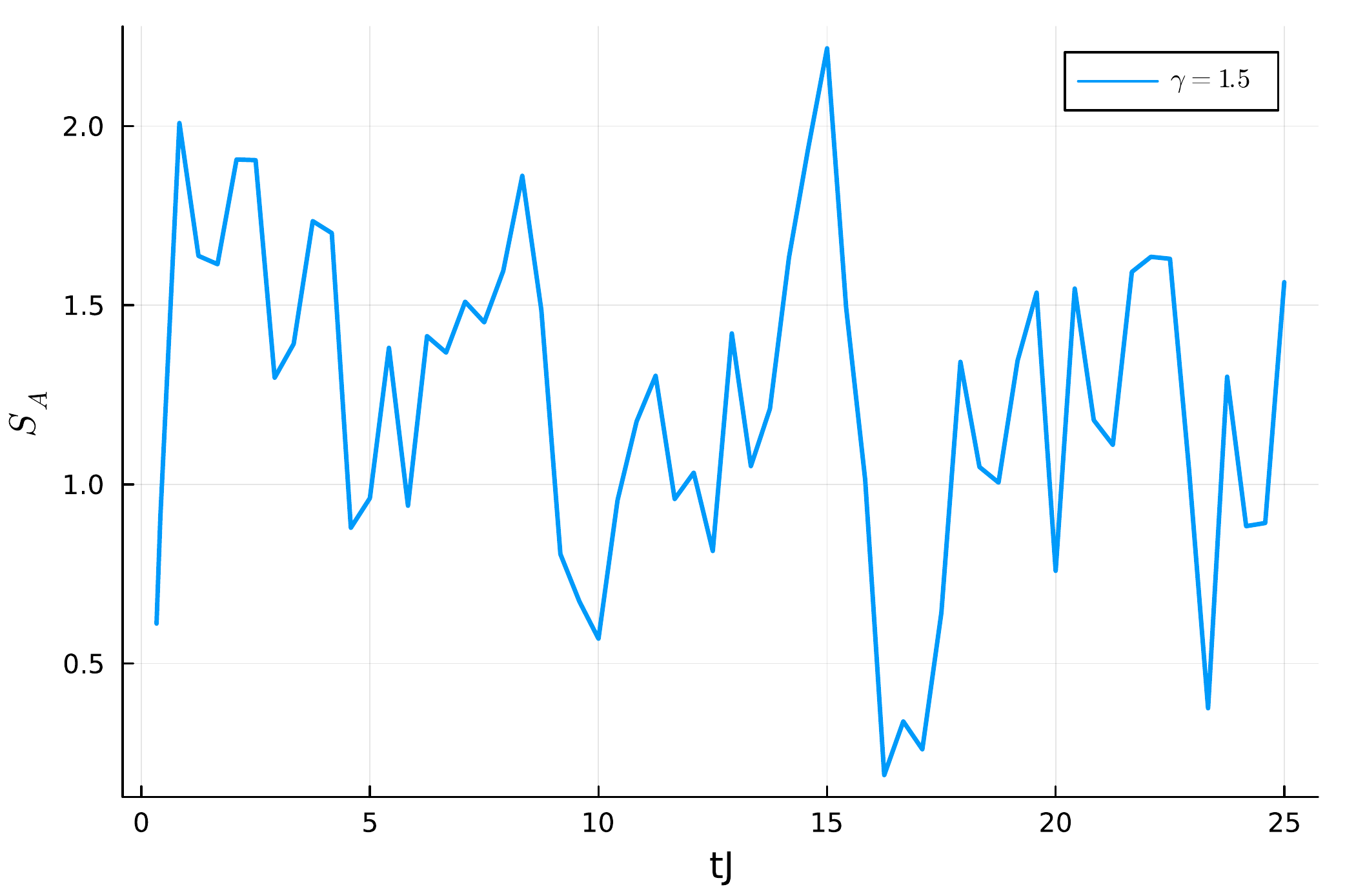} 
	\caption{The entanglement entropy computed on the single realization of a quantum trajectory. Here, $U = J$ and the number of sites $N_s = 6$, and the number of bosons $N_b = 3$, and subsystem size $L_A=3$. The measurement rate is $\gamma/J =1.5$.} 
	\label{single_2}
\end{figure}

Since from the prescription above, the propagation of the initial density matrix $\rho^D(t)=	|\phi^{D}(t)\rangle \langle \phi^{D}(t)|$ in a given time step is:
\begin{equation}
	\begin{aligned}
		\overline{\rho^D(t+\delta t)}=&(1-\delta p)	\frac{|\phi^{(1),D}(t+\delta t) \rangle}{\sqrt{1-\delta p}} \frac{\langle\phi^{(1),D}(t+\delta t)|}{\sqrt{1-\delta p}} +\delta p \sum_a \Pi_a \frac{\hat{L}_{a,L}\hat{L}_{a,R}|\phi^{D}( t)\rangle}{\sqrt{\frac{\delta p_a}{\delta t}}} \frac{\langle\phi^{D}( t)|\hat{L}_{a,R}^{\dagger}\hat{L}_{a,L}^{\dagger}}{\sqrt{\frac{\delta p_a}{\delta t}}}\\
		=&|\phi^{(1),D}(t+\delta t)\rangle \langle \phi^{(1),D}(t+\delta t)|+\eta(t)\delta t \sum_a \frac{\hat{L}_{a,L}\hat{L}_{a,R}|\phi^{D}( t)\rangle \langle\phi^{D}( t)|\hat{L}_{a,R}^{\dagger}\hat{L}_{a,L}^{\dagger}}{\mathrm{Tr}\left(\sum_{b=1}^{n}\hat{L}_{b,L} \hat{L}_{b,R}\rho^{D}(t)\hat{L}_{b,L}^{\dagger}\hat{L}_{b,R}^{\dagger}\right)} \\
		=&-i\left(\hat{H}^D_{\text{eff}}\rho(t)-\rho(t)\hat{H}^{D \dagger}_{\text{eff}}\right)\delta t+\eta(t)\delta t \sum_a \frac{\hat{L}_{a,L}\hat{L}_{a,R}|\phi^{D}( t)\rangle \langle\phi^{D}( t)|\hat{L}_{a,R}^{\dagger}\hat{L}_{a,L}^{\dagger}}{\mathrm{Tr}\left(\sum_{b=1}^{n}\hat{L}_{b,L} \hat{L}_{b,R}\rho^{D}(t)\hat{L}_{b,L}^{\dagger}\hat{L}_{b,R}^{\dagger}\right)}. \\
	\end{aligned}
\end{equation}
Here, the $\overline{\rho^D(t+\delta t)}$ denotes a statistical average over trajectories. It is straightforward to see that the stochastic propagation given by this quantum trajectories method is equivalent to our generalized Lindblad equation Eq.~\eqref{general_density} after taking a stochastic average over trajectories.
\\We use this quantum trajectories method to numerically calculate the half-system entanglement entropy in a larger system size similar to Fig.~\ref{time_evolution} and Fig.~\ref{area_law}. Also, since we are calculating the entropy using the wave function of the double system whose dimension of the Hilbert space is the square of that of the single system, the total single system size that we can simulate efficiently is therefore limited. Here, we calculate the half-system entanglement entropy at total system size $L=2,4,6,8$ and the dissipation strength $\gamma =0.5,1,3,4$ using this quantum trajectories method, and the average longtime saturation entanglement entropy is shown in  Fig.~\ref{saturate_trajectory}.
\\For small values of $\gamma$ ($\gamma/J=0.5$), we see an almost linear growth in the entanglement entropy as the total system size increases, which indicates that there could be a volume-low scaling at a larger system size. In comparison, for large values of $\gamma$ ($\gamma/J=4$), we see a seemly logarithmic growth in the entanglement entropy as the total system size increases, which indicates that there could be a logarithmic critical scaling or an area law (constant) scaling at a larger system size. Also, this result is in line with the understanding that there could be an entanglement phase transition in this process. Also, we illustrate the entanglement entropy computed on a single realization of a quantum trajectory in Fig.~\ref{single_1}. and Fig.~\ref{single_2}. 

The average number of measurements in our simulation is 
\begin{equation}
	N_{mean} =\frac{T_{total}}{dt}L\gamma dt = T_{total}L\gamma.
\end{equation}
For instance, in our simulations, $T_{total}= 30, L = 8,\gamma =1$,  each realization of the stochastic protocol consists of $N_{mean} = 240$ projective measurements on average. Here, we chose a set of $N =100$ quantum trajectories for $L=8$ and $N =1000$ quantum trajectories for $L=2,4,6$.

\section{The entanglement entropy of complement subsystems}
In this section, we will prove that the entanglement entropy calculated from the entropy formula the Eq.~\eqref{SA_definition} for the evolved density matrix the Eq.~\eqref{general_density} is the same regardless of whether one computes partial trace over the subsystem $A$ or subsystem $B$. Here, subsystems $A$ and $B$ are complement subsystems of the total system.
From the Eq.~\eqref{SA_double}, we have
\begin{equation}
	\label{SA_double_short}
	S_A^D =-\log  \left\{ \sum_{b=0}^m \tilde{p_b} \mathrm{Tr}_{A}\left[\rho_{A,b}^2 \right] \right\}.
\end{equation}
Also, if we change the subsystem $A$ to its complement subsystem B, we have
\begin{equation}
	\label{SB_double_short}
	S_B^D =-\log  \left\{ \sum_{b=0}^m \tilde{p_b} \mathrm{Tr}_{B}\left[\rho_{B,b}^2 \right] \right\}.
\end{equation}
with $\tilde{p_b} =\left[\mathrm{Tr}\left(\hat{M}_b\rho_0(t)\hat{M}_b^{\dagger}\right)\right]^{2}$ and $\rho_b=\frac{\hat{M}_{b}\rho_0(t)\hat{M}_{b}^{ \dagger}}{\mathrm{Tr}\left(\hat{M}_b\rho_0(t)\hat{M}_b^{\dagger}\right)}$.
Here, each $\rho_b$ is pure if we start from a pure state. For a pure state density matrix, using the Schmidt decomposition, we can prove that $\mathrm{Tr}_A\left[\rho_{A,b}^2 \right]=\mathrm{Tr}_B\left[\rho_{B,b}^2 \right]$, thus we obtain
\begin{equation}
	\begin{aligned}
		S_B^D &=-\log  \left\{ \sum_{b=0}^m \tilde{p_b} \mathrm{Tr}_{B}\left[\rho_{B,b}^2 \right] \right\}\\
		&=-\log  \left\{ \sum_{b=0}^m \tilde{p_b} \mathrm{Tr}_{A}\left[\rho_{A,b}^2 \right] \right\}\\
		&=S_A^D.
	\end{aligned}
\end{equation}

\section{Numerical results of the entropy computed from the single-copy master equation}
In this section, we will give some numerical results about the system driven by the generalized Lindblad equation the Eq.~\eqref{post_selection_partial} in the original single system, and we set $\eta(t)=\gamma$ as a time-independent measurement rate. Also, we set the projection measurements as
\begin{equation}
	\hat{L}_{i,0}=\frac{1}{\sqrt{L}}(1-\hat{n}_{i}),\  \hat{L}_{i,1}=\frac{1}{\sqrt{L}}\hat{n}_{i}.
\end{equation}
The Hamiltonian of is also the hard-core Bose Hubbard system 
\begin{equation}
	\hat{H}=-J\sum_{\langle i,j\rangle}\hat{b}_{i}^{\dagger}\hat{b}_{j}+U\sum_{\langle i,j\rangle}\hat{n}_{i}\hat{n}_{j}.
\end{equation}
as the section \ref{Numerical_section} for comparison. Here, $J$ is the strength of the nearest neighbor hopping, and $U$ is the strength of the nearest neighbor interaction. 

In our following numerical calculation, we set $J=U=1,N_s=6,N_b=3$. $N_b$ is the total number of the hard-core boson. We denote the left half of the system as subsystem $A$ and the rest of it as subsystem $B$.  We then calculate the entanglement entropy $S_A$ defined in the Eq.~(\ref{SA_definition}). We choose the initial state as the a product state in the particle number basis $|000111\rangle$.
\begin{figure}[h] 
	\centering 
	\includegraphics[width=0.62\textwidth]{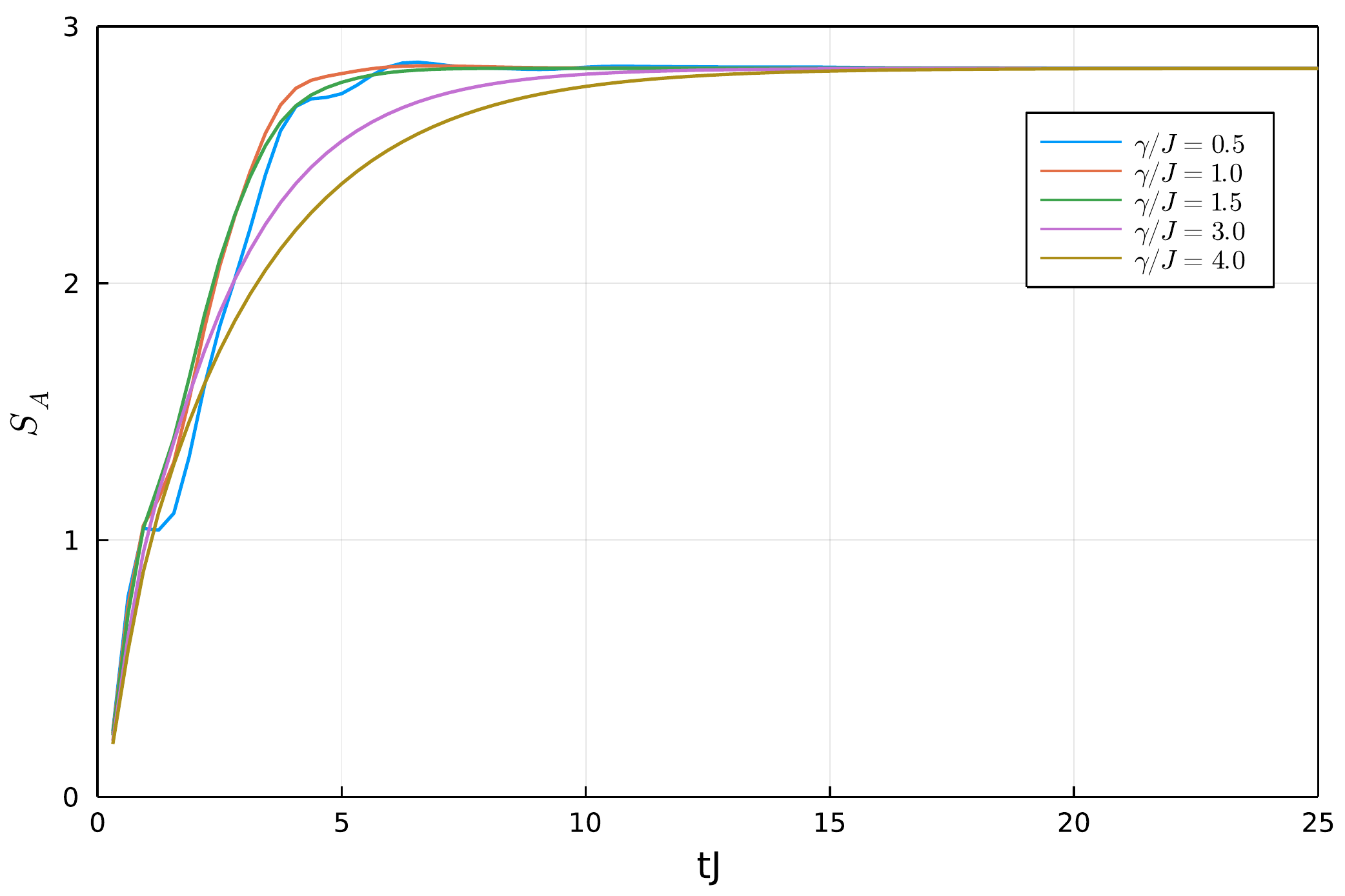} 
	\caption{The dynamics of the entanglement entropy $S_A$ as a
		function of $tJ$. $\gamma$ is the measurement rate. Different curves have different $\gamma$ in the unit of $J$. Here, $U = J$ and the number of sites $N_s = 6$, and the number of bosons $N_b = 3$, and subsystem size $L_A=3$. The density matrix is evolved by the single-copy master equation the Eq.~\eqref{post_selection_partial}.}
	\label{single_time_evolution}
\end{figure}

\begin{figure}[h] 
	\centering 
	\includegraphics[width=0.6\textwidth]{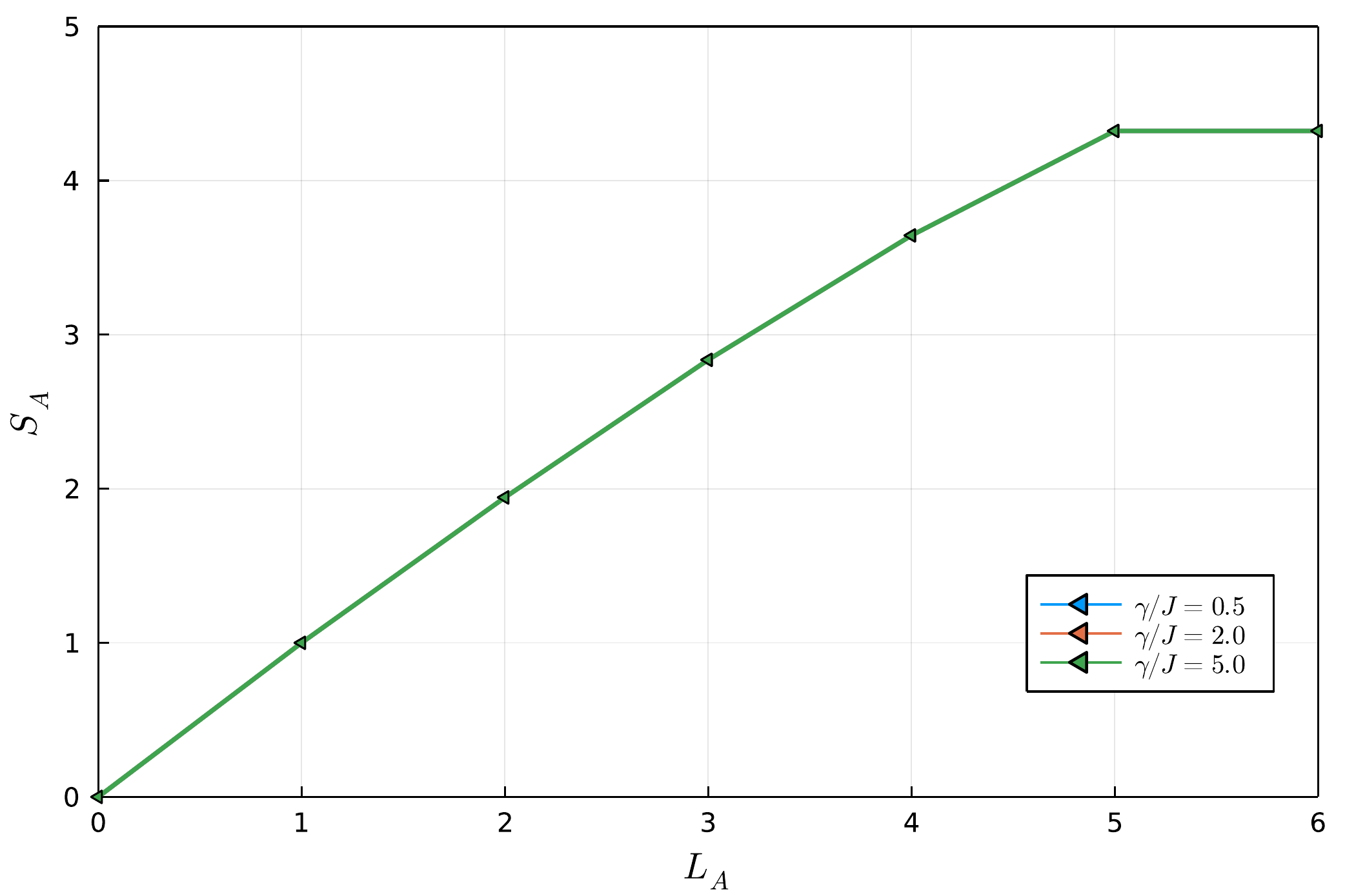} 
	\caption{Saturation value of entanglement entropy $S_A$ of the subsystem $A$ with different subsystem size $L_A$. Different curves have different $\gamma$ in the unit of $J$. Here, $N_s=6$, $N_b=3$, $U/J=1$. The density matrix is evolved by the single-copy master equation the Eq.~\eqref{post_selection_partial}.} 
	\label{single_saturate} 
\end{figure} 
As shown in Fig.~\ref{single_time_evolution}, the entanglement entropy $S_A$ between subsystems $A$ and $B$ increases as expected in a normal chaotic system, and it then saturates to a non-zero value. However, compared with the dynamics of the entanglement entropy $S_A$ in the Fig.~\ref{time_evolution}, it does not decrease in this process, and the saturation value of $S_A$ shown in the Fig.~\ref{single_saturate} is the same as measurement rate changes. 

Therefore, we find that there is no entanglement phase transition in this process, and there is no surprise since we have mentioned in the  section \ref{entanglement_section} that the entanglement entropy calculated from the single-copy master equation is $S^{\text{total}}_{A}$ defined in the Eq.~\eqref{s_total} as
\begin{equation}
	S^{\text{total}}_{A}\equiv-\log\left\{\mathrm{Tr}\left[\left(\sum_{c=1}^m p_c \rho_{A,c}\right)^2\right]\right\} .
\end{equation}
Here, $p_c$ is the probability of getting $\rho_{A,c}$, and it satisfies $\sum_c p_c=1$. Since $S^{\text{total}}_{A}$ also includes the entropy corresponding to the probability distribution of different measurement results, there is no entanglement phase transition regarding this entanglement entropy.

\section{ The detail information about the numerical integration technique used in producing Fig.~\ref{time_evolution} and Fig.~\ref{area_law}}
We used the Runge-Kutta 4th-order (RK4)  method for the approximate solutions of simultaneous nonlinear equations about the double system density matrix. For the Fig.~\ref{time_evolution} and Fig.~\ref{area_law}, we used $N_t = 11000$ time steps.

\section{ The comparison between Eq.(6) with the assumption $\sum_{a=1}^{n}\hat{L}_a^{\dagger}\hat{L}_a=\mathcal{I}$ and the general Lindbald equation.}
\label{difference_section}
In deriving the Linblad-like equation in the section.~(\ref{measurement_Lindblad}), we assume that the measurement is performed on a complete basis, thus the anti-commutator part of the Lindblad master equation is trivial and amounts to an identity matrix. In general, the Lindblad master equation does not need to satisfy the Eq.~(\ref{complete}). Thus, the equation we obtain with the condition Eq.~(\ref{complete}) is a special case of the general Lindblad equation.  
\\The method we have used here to derive this Lindblad-like equation is first evolving the system under unitary evolution and then doing projection according to the probability $p(t)$ on the system. The density matrix after this probabilistic measurement is:
\begin{equation}
	\label{probabilistic_measurement}
	\rho^{M}(t+\delta t)=\left[1-P(t+\delta t)\right]\rho(t+\delta t)+P(t+\delta t)\sum_{a}\hat{L}_a\rho(t+\delta t)\hat{L}_a^{\dagger}.
\end{equation} 
Here, the quantum jump terms $\sum_{a=1}^n\hat{L}_a\rho(t )\hat{L}_a^{\dagger}$ in the equation Eq.~(\ref{without_post}) totally comes from doing measurement on the system. Thus, doing projection on a complete basis directly leads to the condition Eq.~(\ref{complete}).  Also, this probabilistic measurement process in the Eq.~(\ref{probabilistic_measurement}) leads to the anti-commutator part of the Lindblad master equation coming from $P(t+\delta t)\rho(t+\delta t) $. Thus, the anti-commutator part amounts to an identity matrix. Also, if we do not assume this probabilistic measurement process in Eq.~(\ref{probabilistic_measurement}), the system under unitary evolution together with the measurement process can be described in other ways.   For instance, by considering the system being driven by a non-hermitian Hamiltonian\cite{weak_measurement_2021,critical_exponent_Fermion2}, we normalize the quantum state after each step of this protocol. In these protocols,we have $\sum_{a=1}^{n}\hat{L}_a^{\dagger}\hat{L}_a\neq\mathcal{I}$ in general. 

\end{document}